\documentclass{ar2e}
\usepackage{graphicx}
\def\lsim{\mathrel{\rlap{
\lower3pt\hbox{\hskip-3pt$\sim$}}
\raise1pt\hbox{$<$}}}
\newcommand{\gsim}{\raisebox{-0.7ex}{$\stackrel{\textstyle >}{\sim}$ }}
\begin{document}

\jname{Annu. Rev. Nucl. Part. Sci.}
\jyear{2001}
\jvol{51}
\ARinfo{?-?/?/?-01}

\title{NEUTRINO PROPAGATION  IN DENSE ASTROPHYSICAL SYSTEMS  }

\markboth{Prakash, Lattimer, Sawyer, and Volkas }
{Neutrino Propagation in Dense Astrophysical Systems } 
\author{Madappa Prakash and James M. Lattimer
\affiliation{Department of Physics \& Astronomy, 
State University of New York at Stony Brook \\
Stony Brook, NY 11794-3800, U.S.A.}
Raymond F. Sawyer
\affiliation{Department of Physics, 
University of California at Santa Barbara \\ 
Santa Barbara, California 93106, U.S.A.} Raymond R. Volkas
\affiliation{School of Physics\\
Research Center for High Energy Physics, 
The University of Melbourne \\
Victoria 3010, Australia }}
\begin{abstract}
Even the elusive neutrinos are trapped in matter, albeit transiently,
in several astrophysical circumstances.  Their interactions with the
ambient matter not only reveal the properties of such exotic matter
itself, but also shed light on the fundamental properties of the
neutrinos themselves.  The physical sites of interest include the
early universe, supernovae, and newly-born neutron stars.
Detection of neutrinos from these vastly different eras using the new
generation of neutrino detectors holds great promise for enhancing our
understanding of neutrino-matter interactions and astrophysical
phenomena.
\end{abstract}
\maketitle

\section{INTRODUCTION }

In accelerator neutrino physics, a neutrino interacts once, with a
constituent of the matter in a detector, to produce a signal. But in
astrophysical situations the surrounding matter can make a big
difference to the neutrino physics. Some examples are:

\noindent (1) In their passage from the core of the sun to the solar
surface, electron neutrinos ($\nu_e$) feel an ``index of refraction",
or altered energy-momentum connection, that is different from that for
the other neutrino species, 
$\nu_\mu, \nu_\tau$.
 This difference is
induced by forward scattering from electrons in the solar plasma. If
there is in addition a ``mass matrix" that would cause vacuum $\nu$
flavor oscillations, the changing index of refraction seen by a 
$\nu_e$ as it moves into regions of less density can give rise to the MSW
effect. \footnote{A recent review of neutrino oscillation physics is
given in Ref.~\cite{haxton}.}

\noindent (2) The core of a collapsing star, and subsequently the
interior of a newborn neutron star, or proto-neutron star (PNS), formed
during the supernova process, is opaque to neutrinos.  Initially,
because neutrinos are trapped in the PNS,
deleptonization of matter has yet to occur and the total number of
leptons per baryon is about 0.4.  Only after times of tens of seconds
can neutrinos diffuse outwards and escape, but in so doing, they heat
the matter through which they pass in a process reminiscent of Joule
heating.  Following this deleptonization is a cooling epoch, during
which neutrinos continue to transport energy to the star's surface.
The combination of extreme neutrino degeneracy and high temperatures
and densities creates a unique environment in which the detailed
neutrino processes and the resulting transport phenomena play an
essential role.  The escaping neutrino fluxes are crucial for the
supernova dynamics, and are also important in the heavy element
nucleosynthesis that might occur in the ejected envelope of the
formerly collapsing star.  Certainly, they determine the
characteristics of the neutrino pulse that can be observed in
terrestrial detectors \cite{supernova}.

\noindent (3) In the standard model (three $\nu$ flavor) evolution of
the early universe, in the temperature region of, roughly, $0.2 <
k_BT/{\rm MeV} < 10$, the $\nu$'s and $\bar \nu$'s are very nearly in
thermal equilibrium with the other particles in the plasma until the
temperature drops below 1 MeV or so. The corrections to the
distributions near this point of $\nu$ decoupling have been calculated
quite well and provide a small change in the predictions of primordial
$^4$He abundance. But there are models, involving flavor mixing and
the possibility of a fourth neutrino flavor, in which the evolution of
neutrino distributions, including the effects of interactions with the
surrounding plasma, must be studied carefully at higher temperatures
(see Section 4 below).

\subsection{Orders of Magnitude} 
We choose units so that $\hbar=c=k_B=1$.  Then the weak coupling
constant is given by $G_F=1.166\times 10^{-11}~{\rm (MeV)^{-2}}$. Number
densities in these units for some systems of interest, are: 
\newline
Solar center (electrons); $n_e \approx 6 \times 10^{25}~{\rm
cm}^{-3}\approx 5\times 10^{-7}~{\rm (MeV)^{3}}$
\newline 
Earth core
(electrons); $n_e \approx 5\times10^{-6}~{\rm (MeV)^{3}}$
\newline
Supernova-core to neutrinosphere (nucleons) $n_N=[10^4-10^6 ]~{\rm
(MeV)^{3}}$ 
\newline Early universe in the region, 1 $<T/{\rm MeV}<$100;
$n_{\rm particles} \approx 5 \times T^3$. \\

We can estimate the length scales over which the effects of neutrino
interactions with the media can be appreciable in the various cases:

\noindent (a) At solar center densities, the index of refraction 
correction from $\nu_e$-e forward scattering introduces a phase change 
of $2\pi$ in a distance \\ 
$\approx (n_eG_F)^{-1}\approx 4
\times 10^{18}~{\rm (MeV)^{-1}}\approx 2\times 10^7$ cm. 

\noindent (b) The solar center mean free path for $\nu$ scattering is: \\ 
$\lambda_f
\approx (n_eG_F^2 E_\nu^2)^{-1}\approx
2\times10^{28}~{\rm (MeV)^{-1}} 
\approx 4\times 10^{17} ({\rm MeV}/E_\nu)^2$ cm. \\
This is far too long to be of any interest. 

\noindent (c) The supernova center mean free path for neutrino scattering (but
subject to substantial medium-dependent corrections to be discussed in
this review) is \\ 
$\lambda_f \approx (n_NG_F^2 E_\nu^2)^{-1}
\approx 2\times 10^{5} ({\rm MeV}/E_\nu)^2$ cm. 

\noindent (d) For the early universe it is the neutrino collision time 
that is of interest \\
$\lambda_f /c\approx {\rm 10^{-21}}/(G_F^2 T^5)
\approx (T/{\rm MeV})^{-3}$ s. \\ 
This, coupled to the expansion time,
$t$(s)$\approx [T(\rm{MeV})]^{-2}$, shows that $\nu$'s decouple from
the medium at $T\approx 1~{\rm MeV}$.

\subsection{Scope of this Article}

In this review, we shall concentrate on the physics of neutrinos in
supernovae, PNSs and in the early universe.  For these
problems one needs to follow the evolution of neutrino distributions
in time.  In all cases, we envision beginning with an initial
condition with postulated or known neutrino distributions.
In the early universe problem, the initial time can be taken to be
sufficiently early for $\nu$ oscillations not to have affected the
distributions; an initial temperature of tens of MeV suffices in the
models that will be discussed. 

In the supernova and PNS problems, we focus on the
microphysics of neutrino interactions with matter.  For the supernova
case, neutrinos are initially freely streaming and act mainly as a
cooling mechanism.  Shortly after collapse begins, the neutrino mean
free path becomes less than the size of the collapsing core, and
neutrinos become trapped on dynamical time scales.  During this
period, there is a complex interaction between transport and
hydrodynamics.  While we do not discuss supernova simulations in detail, we
summarize the literature of recent results.  For the PNS
problem, which is quasi-hydrostatic, the initial distribution of
neutrinos is that of a highly degenerate Fermi gas.  We present
results of transport simulations in PNSs for a variety
of assumptions regarding the composition of matter.

In all three problems the dynamics of the system provides a time
changing environment. And in these problems the particles other than
$\nu$'s in the soup stay in near thermodynamic equilibrium.
\footnote{In the early universe, as we cool from, say, $T=10$ MeV
to 0.2 MeV the neutrons and protons go far out of chemical
equilibrium, but the nucleons are far too sparse to affect the
neutrino distributions significantly.}

We attempt to pose cleanly some of the questions that have occupied
theorists in these areas, to sketch somewhat qualitatively the methods
that have been used, and to give some examples of results. A prototype
neutrino evolution calculation involves three steps: \\ 
\noindent (1) deriving a
kinetic equation that describes the macroscopic development of the
evolution of the neutrino distribution; \\ 
\noindent (2) determining rate functions
in the kernel of this equation that describe the local scattering,
production and annihilation of the neutrinos; and \\
\noindent (3) solving the equation
for a physical configuration. 

Calculations of phenomena in this area are necessarily approximate but
we shall attempt to define precisely the quantities that enter the
equations.  We must caution the reader that in most of the problems
that we shall discuss there are, at present, few definitive
results. In the first place, to the extent that neutrino mixing (or
``oscillations") are involved, the mixing parameters are still
undetermined. Furthermore, it is essential to follow the time
development of non-equilibrium distributions of several species
simultaneously. Finally, in the supernova and PNS applications,
nuclear forces play an essential role in the opacities, and the
macroscopic environment to be used in the supernova case is dependent
on hydrodynamical calculations as well.

We warn the reader that the references are not exhaustive.

\section{NEUTRINO  MIXING AND INDEX OF REFRACTION EFFECT} 

The basis for what follows will be the standard model of neutrino
interactions, in which all coupling is to the left handed neutrino
current operators, in places supplemented by neutrino mass and mixing
effects, or by the inclusion of a "sterile" neutrino, which has no
interactions outside of the mixing term.  If there is neutrino mass
and mixing the standard model is augmented with a term of the form $
H_{mix}=\sum_{\alpha,\beta}m_{\alpha,\beta}\bar \psi_\alpha
\psi_\beta$, where the $\psi$'s are the $\nu$ fields and
$(\alpha,\beta)$ are flavor indices, taking the values $e, \mu, \tau$
and, in the case of the consideration of a sterile neutrino, $s$.

The differences in the forward scattering amplitude of the different
flavors of $\nu$'s in the various media, as calculated in first order
in $G_F$, give rise to a relative energy shift matrix $\Delta_{\alpha
,\beta}$ that is first order in $G_F$, independent of the neutrino
energy, diagonal in the flavor indices, and proportional to the
density of scatterers in the medium. In the usual MSW considerations
this comes from the electron density, $n_e$, only and is given by
$\Delta_{\alpha,\beta}=\delta_{\alpha,e}\delta_{\beta,e}\sqrt{2} G_F
n_e$.  However, in the interior of a supernova there may be enough
$\mu^-$ present briefly to make a significant contribution to the
forward scattering differences. Sterile neutrinos would add to the
matrix, $\Delta$, as would one loop weak corrections under some
circumstances \cite{raff1}.

In all cases of interest, we have $m,\Delta\ll E_\nu$. Then standard
Dirac technology combines the neutrino mass and index of refraction
effects in an effective Hamiltonian for the $\nu,\bar\nu$ system,
not including interactions, except for forward scattering,
\begin{equation}
H^{(\nu)} = \sum_{{\bf p},\alpha,\beta}\Bigr\{a^{\dagger}_\alpha
({\bf p})a_\beta({\bf p})
\Bigr[E\delta_{\alpha,\beta}+\lambda^{(+)}_{\alpha,\beta}(E)\Bigr]
+b^{\dagger}_\alpha({\bf p})b_\beta({\bf p})
\Bigr[E\delta_{\alpha,\beta}+\lambda^{(-)}_{\alpha,\beta}(E)
\Bigr]\Bigr\} \,,
\label{1}
\end{equation}
where $E=|\bf p|$ and the $\lambda$ matrices are given by $\lambda
^{(\pm)}(E)=m^2 / (2E) \pm\Delta$. Here $a_\alpha ({\bf p})$ and
$b_\alpha ({\bf p})$ are the respective annihilation operators for
left-handed neutrinos and antineutrinos of flavor $\alpha$. The small
admixtures of right handed $\nu$'s created by the mass term, for the
case of Dirac neutrinos, are inconsequential in our applications. The
magnitudes of the elements of the mass$^2$ matrix that have been
suggested in the literature range from ${ \rm 10^{ -15} (eV)^2 }$ to
${ \rm 1}$(eV)$^2$ depending on the application.

\section{KINETIC EQUATIONS}

If there are no $\nu$ oscillations, the kinetic equations will be for
distribution functions, $\rho_\alpha ({\bf p,r, \rm t})$ for each
flavor $\alpha$ of $\nu$ and $\bar \nu$, where the $r$ dependence is
macroscopic in scale and originates from spatial variation of the
properties of the medium, or in the initial condition taken for the
neutrino distribution. For the case with neutrino oscillations,
however, there can be off-diagonal flavor coherence that extends over
large distances, and the system is no longer described in terms of a
set of probabilities for the occupancy of the modes of the neutrino
fields for each flavor. Instead we introduce a $\nu$ momentum-flavor
density matrix, and discuss its evolution.

At $t=0$ we take the density matrix for the complete system to be of
the form of a product of a neutrino matrix times an everything-else
matrix, the everything-else being in thermal equilibrium; of course,
these become entangled at later times. Introducing the operator,
$\Lambda({\bf k},t=0)_{\alpha,\beta}=a^\dagger ({\bf k})_\alpha a({\bf
k})_\beta$, we define the momentum-flavor density matrix,
\begin{equation}
\rho_{\alpha,\beta}({\bf k},t)=Z^{-1}_{\ne \nu}~{\rm Tr }
\Bigr [ e^{-( H_{\ne \nu}-\sum_{i \ne \nu}\mu_i N_i)/T}
\sum_i  w_i \langle i \,\, ,\nu's |\Lambda({\bf  k},t)_{\alpha,\beta}|i \,\, ,
\nu 's\rangle\Bigr ] \,,
\label{2}
\end{equation}
where $\Lambda ({\bf k}, t)=\exp(iHt)\Lambda ({\bf k},0) \exp(-iHt)$ ,
$H_{\ne\nu}$ is the Hamiltonian with all terms containing the $\nu$
coordinates left out, and $Z_{\ne \nu}$ is the partition function
associated with this Hamiltonian. The states, $|i\, ,\nu's\rangle$,
represent all possible multi-$\nu$ states, indexed with $i$. The
coefficients $w_i$ in the inner sum might be chosen in ways that are
far from a thermal equilibrium configuration (for example, such that
there are initially no sterile $\nu$'s, in models that contain a
sterile $\nu$). $H$ is the complete Hamiltonian.  We take the neutrino
interaction with other particles to be of the
form\footnote{Generalizations to a sum of terms of this form are
immediate, but give lengthier expressions in the evolution equations.}
\begin{equation}
H^{(\nu)}_{int}=g\sum_{\alpha,\beta} \zeta_{\alpha,\beta}\bar\psi_\alpha 
[\gamma_\mu.....]\psi_\beta \times[{\rm other\, fields}] \,,
\label{3} 
\end{equation}
where the matrix $\zeta$ acts in flavor space, and is generally
diagonal in flavor. For example, if we are discussing active-sterile
$\nu$ mixing we have, $\zeta_{a,a}=1$,
$\zeta_{s,s}=\zeta_{a,s}=\zeta_{s,a}=0$. For $\nu_e$ mixing with an
active species, $a$, in an environment containing electrons and
positrons but no other leptons, we have, $\zeta_{e,e}=1+c_e$,
$\zeta_{a,a}=1$, where $c_e$ gives the effect of the charged current
e, $\nu_e$ scattering term.  There is no exact equation governing the
time evolution of the function, $\rho_{\alpha,\beta}({\bf k},t)$. Indeed,
the evaluation of the two neutrino fields in $\Lambda({\bf k})$ at the
same momentum $\bf k$ rules out the derivation of such an equation. But if
we 
had defined a density matrix with momentum indices, $\bf k,k'$ then,
at times large compared to a typical $E_\nu ^{-1}$ we would find that,
for all practical purposes, this density matrix was very nearly
diagonal in these indices. We can qualitatively describe the
simplification as due to the rapid decoherence between the parts of
the $\nu$ wave function that are scattered at appreciably different
momentum transfer (and/or energy transfer) from components of the
medium at different (randomly situated) points in space, the density
matrix in momentum space becoming nearly diagonal so fast that we need
to consider only the diagonal elements in the $\bf p$ space. Finally,
we consider only systems in which the local averages of the
densities, compositions, and temperatures varies slowly in space,
compared to all quantum mechanical distances except for the neutrino
oscillation lengths. The evolution equation is then,
\begin{eqnarray}
{\partial\over \partial t}\rho({\bf k},{\bf r},t)+|{\bf k }|^{-1}
{\bf k \cdot \nabla}\rho({\bf k},{\bf r},t) = 
-i [\lambda(E),P({\bf k},{\bf r},t)]+ 
\nonumber\\
+{1\over 2}\sum_{\bf k_1}\Bigr [\zeta \rho({\bf k_1},{\bf r},t)\zeta 
[1-\rho({\bf k},{\bf r},t)]+[1-\rho({\bf k},{\bf r},t)]
\zeta \rho({\bf k_1},{\bf r},t)\zeta \Bigr ]\Gamma({\bf k}_1,{\bf k})
\nonumber\\
-{1\over 2}\sum_{\bf k_1}\Bigr[\zeta [1-\rho({\bf k_1},{\bf r},t)]\zeta
\rho({\bf k},{\bf r},t)+\rho({\bf k},{\bf r},t) 
\zeta[1-\rho({\bf k_1},{\bf r},t)]
\zeta \Bigr] \Gamma({\bf k},{\bf k}_1) \,.
\label{4}
\end{eqnarray}
The function, $\Gamma({\bf k,\bf k_1})$, is the differential rate at
which neutrinos would scatter from energy $\bf k$ to momentum ${\bf
k_1}$ if we took $\zeta=1$, that is, if the neutrino interaction were
flavor independent. \footnote{As a concrete example, in the case in
which one of the neutrinos is an electron neutrino, the plasma is
constituted of electrons and positrons, and we consider $\nu_e$,
$\nu_\mu$ mixing, $\Gamma$ stands for rate of scattering of $\nu_\mu$,
and $\zeta$ is given as above. The squared amplitude contributions
from scattering from electrons get added to those from positrons. So
we can use $\zeta_{ee}=1+c$, just as though there were a single term
in equation~(\ref{3}).}

Many alternative paths to a derivation of equations like (\ref{4}) are
found in the literature.  If we consider only the terms that are
linear in $\rho$, as for the case of a single neutrino traversing the
medium, then with the right identifications equation~(\ref{5}) can be
seen to be the same as the ``quantum kinetic equation" of McKellar and
Thomson \cite{MandT}, taken in the same limits. Using a somewhat
idealized form of interaction, Raffelt, Sigl, and Stodolsky
\cite{raff2} have given an elegant derivation that includes Fermi
statistics (i.e. the nonlinear terms on the RHS); they call the result
the ``non-Abelian Boltzmann equation." Again for the energy conserving
case, but without Fermi statistics, Loreti and Balantekin \cite{baha}
have used the equation, which they describe as a variant of the
``Redfield equation" to analyze MSW transitions in media with
irregular electron density variations in space. In the vast literature
from the condensed matter and quantum optics side, such equations are
often called ``Bloch equations." Typical derivations are given in
Refs. \cite{vankampen,walls}.  A simple derivation of
equation~(\ref{4}) that captures most of the essential aspects of the
physics, for the case of the density matrix of a single neutrino, is
given in Ref.~\cite{bvs}.

A simple application of equation~(\ref{4}) is to active-sterile
mixing, involving two neutrino states that mix, only one of which
scatters. For a translational and rotationally invariant system, we
calculate a density matrix $\rho(E,t)$. We keep only the linear terms
for illustration, and we adopt the representation $ \rho=
{1\over2}(P_0+{\bf P} \cdot \sigma)$, $\lambda= {\bf V}\cdot \sigma$
. The projection operator on the active state is
$\zeta=(1+\sigma_z)/2$, where $\sigma$'s are the Pauli
matrices. Substituting in equation~(\ref{4}) gives
\begin{equation}
{\partial\over\partial t} {\bf P} (E,t)={\bf V}(E)\times {\bf P}(E)
-{1 \over 2} \Bigr[\hat x P_x (E) 
+\hat y P_y(E) \Bigr]
\sum_{E_1}\Gamma(E,E_1)
+\hat z{\partial\over\partial t}P_0 (E,t)
\label{5}
\end{equation}
and 
\begin{equation}
{\partial\over\partial t}P_0 (E,t)=-{1 \over 2}[P_z(E)+P_0(E)]
\sum_{E_1}\Gamma(E,E_1)
+{1 \over 2} \sum_{E_1}\Gamma(E_1,E)[P_z(E_1)+P_0(E_1)] \,.
\label{6}
\end{equation}
For later use in Section (4), it is helpful to define a decoherence
function through $ D \equiv (\Sigma_{E_1} \Gamma(E,E_1))/2 $ and a
repopulation function through $ R \equiv \partial P_0/\partial t $.

\subsection{Extensions of the Evolution Equation} 

In the kinetic equation~(\ref{4}) the sources of change of the
neutrino distribution are transport, flavor precession, and
scattering. We sketch briefly the other effects that should be
included. For economy in notation we can consider the spatially
uniform case, in which the $\nu$ density matrix is given by
$P_{\alpha,\beta}(E)$, as in equation~(\ref{5}).

\paragraph{A. $\nu\bar\nu$ Pair Production and Annihilation:}
These processes occur due to the conversion of charged
lepton-anti-lepton pairs, or from $\nu$ pair bremsstrahlung
processes. To include these, we can introduce a set of densities for
antineutrinos and incorporate the coupling between the two sectors
cominmg from pair processes. However, it is neater, to define the
function $[1-\rho(-E)]$ as the $\bar\nu$ density matrix, to use the
extension of $\lambda(E)$ to negative energies provided by the
definition in equation~(\ref{2}), and to extend the functions, $\Gamma
(E,E_1)$, so that for $ E<0 , E_1>0$ it is the pair production rate,
for $E>0,E_1<0$ the pair annihilation rate, and for $E<0,E_1<0$ the
anti-particle scattering rate, but from energy, $-E_1$, to energy
$-E$. With these conventions the only change to be made in
equation~(\ref{5}), to accomodate pair creation and annihilation will
be to relabel the energy sums in equation~(\ref{5}) to be
$\sum_{E_1=-\infty}^\infty$. In the case $\lambda=0$, $\zeta =1$, the
system would move toward a thermal equilibrium described by a $\nu,
\bar \nu$ Fermi distributions with a single chemical potential
determined by the excess lepton number for each flavor.

\paragraph{B. Expansion of the Universe:}
This is taken into account by using the
time dependent temperature in the statistical factor and adding to the
LHS of equation~(\ref{5}) the term $-[\dot a(t)/a(t)]E(\partial /\partial
E)\rho(E,t)$ where $a(t)$ is the scale factor.

\paragraph{C. Charged-Current Interactions of Electrons with Nucleons:}
This requires an additional term on the RHS of equation~(\ref{5}), 
of the form, \\
$-\delta_{\alpha,e}\delta_{\beta,e}[\rho_{\alpha,\beta}(E)\Gamma_a(E)+
(1-\rho_{\alpha,\beta})\Gamma_e(E)]$,
where $\Gamma_a$ and $\Gamma_e$ are the appropriate differential rates
for absorption and emission, from the medium, of a $\nu_e$.

\paragraph {D. Neutrino-Neutrino Scattering:} 
At the present stage of the development of the formalism the effects
of $\nu-\nu$ scattering must be put in by hand in a way that suits the
application.

In modern supernova calculations in the absence of neutrino mixing, it
is usual to bypass the explicit Boltzmann equation for the evolution
of the density functions, and to do neutrino transport by following
the neutrino distributions numerically, energy bin by energy bin, and
region by region in space. The basic input is the differential rate
function, $\Gamma(E,E_1)$ introduced above. Much of the review to
follow will be concerned with the calculation of these functions in
the presence of strong interactions among the constituents of the
matter.

However, we note that the equation governing the evolution of the
distribution function in the presence of mixing cannot be interpreted
as assigning probabilities, effective immediately, at the outcome of
every neutrino interaction. It is clear that the numerical simulation
of a case with neutrino mixing, involving some phase information that
may propagate through many scattering interactions, will be more
complex, and will require going back to the equation for the
distribution function that is shown above.

\subsection{Determination of the Rate Functions}

In applications we need only terms of order $G_F^2$. The rates, in
media that are in thermal equilibrium are given by thermal averages of
Heisenberg picture current operators. For neutral current scattering
we define a current correlator, $W_{\mu\nu}$,
\begin{equation}
W_{\mu\nu}(q,\omega)=Z^{-1}\int d^4x \;e^{-i \bf q
\cdot \bf x}\rm\it e^{i\omega t}~{\rm Tr}[e^{-\beta( H-\Sigma \mu_iN_i)}j_\mu 
(\bf x \rm\it , t)j_\nu(0,0)]\,,
\label{neutralcorrelator}
\end{equation}
where the $j_\mu$ is the neutral current operator, which we have taken
to be independent of $\nu$ flavor .\footnote{In applications in which
($\nu_e,e$) scattering plays a role, it is convenient to include the
charged current contribution to the ($\nu_e,e$) interaction in this
term, giving a dependence on the $\nu$ flavor indices that we do not
show explicitly.} The quantities $H$ and $N_i$ contain only the
coordinates of the species that are in equilibrium. The differential
rate for neutrino scattering, at angle $\theta$ and with energy loss
of $\omega$, or gain of ($-\omega$), is
\begin{equation}
\frac{d^2 \Gamma}{d \omega \hspace{1 pt} d \cos\theta}=
(4 \pi^2)^{-1}G_F^2 (E_1-\omega)^2[1-f_{\nu}(E_1-\omega)]
\Lambda^{\mu\nu}(q,\omega) W_{\mu\nu}({\bf}q,\omega) \,,
\label{formalneutral}
\end{equation}
where $\bf q =p_1-p_2$ is the momentum transferred from the neutrino
to the medium, $Z$ is the partition function, and
\begin{equation}
\Lambda^{\mu\nu}=(4E_1E_2)^{-1}~{\rm Tr} [p_1 \hspace{-9.5pt}/ 
\hspace{8 pt}(1-\gamma_5)
\gamma^{\mu}p_2 \hspace{-9.5 pt}/\hspace{8pt}\gamma^{\nu}(1-\gamma_5)].
\label{9}
\end{equation}
For the case of charged current reactions there is a parallel
construction, where we now take $j_\mu$ to be the hadronic current
that is coupled to electron emission, with its Hermitean conjugate the
current that is coupled to electron absorption: 
\begin{equation}
W^{(ch)}_{\mu,\nu}( q,\omega)=-iZ^{-1}\int d^4x \; e^{-i {\bf q}
\cdot {\bf x}}e^{i(\omega+\hat\mu) t}
~{\rm Tr} \bigl[e^{-\beta( H-\Sigma \mu_i N_i)}j_{\mu}({\bf x}  , t) 
j^{\dagger}_{\nu}(0,0)] \,,
\label{chargedcorrelator}
\end{equation}
where $\hat\mu$ is the chemical potential difference
$\hat\mu=\mu_n-\mu_p$.  In the early universe application, where the
medium is comprised almost entirely of e$^{\pm}$, $\nu,\bar \nu $, it
is quite a good approximation to replace the formal expressions for
the functions that determine the rates, such as $W_{\mu\nu}$ in
equation~(\ref{formalneutral}), by their vacuum Born approximations, modified
only by the statistical factors for the external particles. But there
are corrections of order 1\% in the predicted $^4$He abundance that come
from careful nonequilibrium calculation of the functions related to
$W_{\mu\nu}$ in the thermal environment \cite{turner,
brownandsawyer}.

In the supernova environment, however, the densities and temperatures
from the core to the neutrinosphere, at the average radius of the last
$\nu$ scattering, are such that the wavelength of a thermal neutrino
is greater than the interparticle spacing, so that neutrino scattering
is a collective phenomenon. In this case a treatment must begin
directly from the dynamical functions defined in equations~
(\ref{neutralcorrelator}) and (\ref{chargedcorrelator}) rather than
from the (cross section$\times$density) expression that applies to
dilute systems or for more energetic neutrinos.  In the center of the
region the densities are on the order of nuclear densities
($n_o\simeq0.16$ fm$^{-3}$) and the matter is quite degenerate, much
like the nuclear matter that has been discussed with respect to the
interiors of heavy nuclei or neutron stars. But nearly all the way to
the $\nu$-sphere, at a density $\approx0.01n_0$ and
with $T \approx$5 MeV, strong interactions among the nucleons will
significantly affect the rates of neutrino processes, due to the fact
that $T$ is less than the per-nucleon interaction energy throughout
most of the region.

\subsubsection{Long-Wavelength Limits}

The results of the detailed calculations that will be summarized below
indicate that the main features of the combined effects of Fermi
statistics and of the nuclear interactions can be described by some
simple limits. When the neutrino has energy of the order $T \ll M_N$
there is relatively small energy transferable to the nucleons. Then
integrating equation~(\ref{formalneutral}) over $\omega$ to get the rate that
is differential in angle only, the correlator $W_{\mu\nu}({\bf
q},\omega)$ has a peak around $\omega=0$ and the multiplying factors
can be evaluated at $\omega=0$. Thus what enters the total rate is the
the energy integral of the correlator itself, or the $t=0$ value of
the time dependent correlator. As illustration we take the vector
current only, with a coupling to a single kind of nucleon, and define
a structure factor, $S(q)$, by
\begin{equation}
n S(q)=(2\pi)^{-1}\int _{-\infty}^{\infty}d \omega ~W_{00}(q,\omega) \,.
\end{equation}
Looking first at a free nearly-degenerate Fermi gas, $T \ll E_F$, we
find that $S(q)$ is less than unity because of the Pauli principle
reduction of phase space. There is some phase space available for two
reasons: \\ 
\noindent (a) The Fermi surface is diffuse, since $T\ne 0$, with some
states below the Fermi energy unoccupied, and some above occupied. No
matter how small the momentum transfers, this gives some room for
transitions, with a rate contribution proportional to $T$. \\ 
\noindent (b) There is
also a $T=0$ limit, but here the neutrino scattering, with momentum
transfer $q$, must excite a state from below to above the Fermi
level. 

At low temperatures these corrections are additive. But when $q
\approx T$ the (a) terms, proportional to $T$, will always predominate
over the (b) terms, by the ratio of the speed of light to the Fermi
velocity \cite{oldestsawyer}. 
(The nucleons are quite non-relativistic in most cases.)

Turning to the case with nucleon-nucleon interactions, again for
thermal q, so that the terms proportional to $T$ dominate, we can use a
powerful classical result: in our one species gas the long wavelength
limit of the structure factor is given by
\begin{equation}
{\rm Lim}_{q \rightarrow 0}~S(q)=n T\kappa_T^{-1} \,.
\label{isotherm}
\end{equation}
Here $\kappa_T$ is the isothermal bulk modulus, $\kappa_T=n[\partial
P/\partial n]_T=n\partial^2 F/\partial n^2$, where $P$ is the pressure
and $F$ is the Helmholtz free energy density. Note that if the
equation of state (EOS) were that of a free Boltzmann gas $P=nT$,
equation~(\ref{isotherm}) gives the uncorrelated value $S=1$. If the
EOS is that of a degenerate Fermi gas, then we get the same answer as
in the dominating term proportional to $T$ in the direct Pauli
blocking calculation reviewed above. Using equation~(\ref{isotherm})
we have replaced a correlation function calculation by an EOS
calculation. This pays dividends for the case of the strongly
interacting gas. For example, in the nearly degenerate case we get the
dominating term in the structure function, proportional to $T$, from
the zero-temperature EOS, for which nuclear matter calculations
provide data. We also get a qualitative lesson directly from
equation~(\ref{isotherm}), namely that attractive interactions, which
soften the equation of state (reduce $\kappa_T$), increase neutrino
scattering while repulsive interactions decrease neutrino scattering.

For the neutrino scattering case, the axial current (Gamow-Teller or
GT for short) terms coming from equation~(\ref{neutralcorrelator})
depend on the nucleon spin density correlators, rather then the
density correlators. These terms provide roughly 3/4 of the
scattering, in the free particles case. It is straightforward to
generalize the above small $q$ analysis. The results are that the GT
terms in the total opacity have a structure similar to
equation~(\ref{isotherm}), but with $\kappa_T$ in the denominator
replaced by $\kappa_{\rm spin}=n_s\partial^2 F/\partial n_s^2$, where
$n_s$ is the spin density of the system. That is to say, in order to
determine the spin density correlators that determine the rate
functions we must calculate the free energy (or energy, in the case of
the nearly degenerate system) with a constraint of non-vanishing
expectation of nuclear spin, to second order in the spin excess. This
would seem to be within the range of variational nuclear matter
calculations \cite{pandharipande}, but it has not been carried through
for the phenomenological potentials that are in use. For the case of
the nuclear matter in nuclei however, to the extent that the term
``nuclear matter" is applicable, it is the modulus $\kappa_{\rm spin}$
that determines the parameters of the giant GT resonances, so that
there is indeed the possibility of using values determined fairly
directly from data rather than those coming from nuclear matter
calculations based on a specific potential.

\subsubsection{Strong and Electromagnetic Correlations}

With these remarks as orientation, we describe the ring graph
calculation of the correlators. Although we will quote results from
more than one source, and more than one exact set of assumptions, the
basic inputs are Fermi liquid parameters that incorporate a good deal
of nuclear phenomenology. These parameters are effectively the
coupling constants of zero range N-N potentials that describe the low
energy excitations of the matter. The ring graph sum has two
properties that are important to us in the light of the above
discussion:

\noindent (1) In the $q=0$ limit, the sum of the ring graphs for the
correlators gives back the connection equation~(\ref{isotherm}) when we take
the system's free energy density to be given by its Hartree
value. There is a similar connection for the spin response part.

\noindent (2) In contrast to the model of a free nucleon gas, in which
the neutrino-nucleon collisions are nearly elastic, the ring results
show that significant gains and losses of neutrino energy may be
realized in the collision. Much of this inelasticity is found to be in
the form of the emission and absorption of spin waves in the medium:
these are the continuous matter analogues of the giant GT resonances
in nuclei. The giant GT resonance was a principal ingredient in the
phenomenological determination of the ``spin-isospin" Fermi liquid
parameter that goes into our model.  The dispersion relation that
describes the nuclear GT resonances (with ``momentum" $\approx
\pi/${\rm nuclear~ radius}) is consistent with the Cerenkov angle
found in \cite{IP,rfs,redd98,redd99a,burr98} for the emission of a
spin wave into the medium.  Whatever residual deficiencies there may
be in the method, or in our underlying simplification of the nuclear
interactions, what follows captures some of the essential physics. The
ring equations, which we shall give in rather schematic form, begin
with the definition of a ``polarization" $\Pi_{\mu\nu}$ through the
following replacements on the LHS and RHS of
equation~(\ref{formalneutral}) $W_{\mu\nu}({\bf q},\omega)\rightarrow
\Pi_{\mu\nu}({\bf q},\omega)$ and $j_{\mu}(\bf x \rm\it ,
t)j_{\nu}(0,0)\rightarrow [j_{\mu}(\bf x \rm\it ,
t),j_{\nu}(0,0)]\theta (t)$, the retarded commutator being the
structure of choice for many-body graph summing.  Then the rate
function $W_{\mu\nu}$ is recaptured through,
\begin{equation}
W_{\mu\nu}(q,\omega)=2 (1-e^{-\beta\omega})^{-1}{\rm Im}
[\Pi_{\mu\nu}(q,\omega) ].
\end{equation}
The function $\Pi_{\mu\nu}(q,\omega)$ can be decomposed in scalar
functions. For simplicity we illustrate the ring approximation
(sometimes referred to as the RPA) for the case of the vector current
part of the neutral current coupling to a single species of nucleon
(or, equivalently, to the isoscalar part of the coupling to protons
and neutrons). In this case the time components $\Pi_{00}(q,\omega)
\equiv \Pi(q,\omega)$ obey the ring equation,
\begin{equation}
\Pi(q,\omega)=\frac{\Pi^{(0)}(q,\omega)}{1-v(q) \Pi^{(0)}(q,\omega)}\,,
\label{a23}
\end{equation}
where $v(q) =\int d^3 x~\exp [i {\bf q \cdot x}]V(x)$ and
$\Pi^{(0)}(q,\omega)$ is the free polarization part.  If we take only
the numerator in equation~(\ref{a23}), we recover the effects of Pauli blocking
alone.

This approach can be fleshed out to include spin and isospin, giving
rise to multichannel equations; then the minimal $v(q)$ in equation 
(\ref{a23}), now a matrix in channel space, can be taken from the
Fermi liquid parameters as estimated, e.g., in \cite{fermiliquid}.
Details and further discussion of these methods can be found in Refs. 
\cite{IP,rfs,redd98,redd99a,burr98}.  

The extension of these methods to encompass a relativistic
field-theoretic description of the matter in which the neutrinos
propogate is straightforward \cite{redd98,redd99a,horo91}; in this
approach the target particle time ordered or causal polarization
tensor is calculated using
\begin{equation}
\Pi_{\mu\nu}=-i \int
\frac{d^4p}{(2\pi)^4} {\rm Tr}~[T(G_2(p)J_{\mu} G_4(p+q)J_{\nu})]\,.
\end{equation}
The Greens' functions $G_i(p)$
(the index $i$ labels particle species) describe the propagation of
baryons at finite density and temperature.  The current operator
$J_{\mu}$ is $\gamma_{\mu}$ for the vector current and
$\gamma_{\mu}\gamma_5$ for the axial current.  Effects of strong and
electromagnetic correlations may be calculated by utilizing the RPA
polarization tensor 
\begin{equation} 
\Pi^{RPA} = \Pi + \Pi^{RPA} D \Pi~, 
\end{equation}
where $D$ denotes the interaction matrix (see \cite{redd99a,horo91} for more
details).

\subsubsection{Neutrino Scattering in Heterogeneous Media}

The coherent scattering of neutrinos from heterogeneous media was
first discussed by Freedman \cite{Freedman} in conjunction with heavy
nuclei in the subnuclear density regime.  For neutrino de Broglie
wavelengths $\lambda_\nu>>R_N$, where $R_N$ is nuclear radius, the
opacity is enhanced by a factor $N^2/6A\approx 100$ compared to that
from the same mass density of nucleons alone~\cite{Freedman}.  In the
supernova environment, coherent scattering dominates all other opacity
sources.  However, nuclei are relatively close together and the
neutrino wavelengths are not necessarily large.  Three important
corrections~\cite{BURLAT} that must be applied which reduce the
coherent cross section include the nuclear structure factor (when
$\lambda_\nu<R_N$), liquid structure effects (when
$\lambda_\nu>R_N~(n_0/n)^{1/3}$, the internuclear spacing), and
polarization of the medium by the electrons (when $\lambda_\nu>R_D$,
where $R_D$ is the Debye radius).  Obviously, effects that influence
the nuclear radius, such as the internuclear spacing, the nuclear
surface energy, and finite temperatures, are very important, and are
equation of state sensitive.  In addition, around $n_0/3$, where the
nuclear phase might transform into an inside-out, or bubble, phase,
structural effects could play a role in neutrino cross sections.

More recently, coherent scattering from droplets of exotic matter (kaon
condensates or quark matter) in the supranuclear density
regime has been also considered \cite{RBP00} for the PNS
environment.

The Lagrangian that describes the neutral current coupling of
neutrinos to the droplet (either a nucleus or a droplet of exotic matter) is
\begin{equation}
{\mathcal{L}}_W = \frac{G_F}{2\sqrt{2}} ~\bar{\nu}\gamma_\mu(1-\gamma_5) \nu
~J^{\mu}_D \,,
\label{nuD}
\end{equation}
where $J^{\mu}_D$ is the neutral current carried by the droplet.  For
non-relativistic droplets, $J^{\mu}_D = \rho_W(x)~\delta^{\mu 0}$ has
only a time like component. Here, $\rho_W(x)$ is the excess weak
charge density in the droplet. The total weak charge enclosed in a
droplet of radius $R_N$ is $N_W=\int_0^{R_N} d^3x ~\rho_W(x)$ and the
form factor is $F(q)=(1/N_W)\int_0^{R_N} d^3x ~\rho_W(x)~
\sin{qx}/qx$. The differential cross section for neutrinos scattering
from an isolated droplet is then
\begin{equation}
\frac{d\sigma}{d\cos{\theta}}= \frac{E_\nu^2}{16\pi}
G_F^2 N^2_W(1+\cos{\theta}) F^2(q) \,,
\label{diff}
\end{equation}
where $E_\nu$ is the neutrino energy, $\theta$ is the
scattering angle, and $q=\sqrt{2}E_\nu(1-\cos{\theta})$ is 
the momentum transfer.  Since the droplets are massive, elastic scattering 
dominates.

The droplet radius $R_N$ and the inter-droplet
spacing are determined by the balance between surface and Coulomb
energies.  In the Wigner-Seitz approximation, the cell radius is
$R_W=(3/4\pi N_D)^{1/3}$, where the droplet density is $N_D$. Except
for one aspect, we will neglect coherent scattering from more than one
droplet. If the droplets form a lattice, Bragg scattering will
dominate and our description would not be valid. But for low density
and a liquid phase, interference from multiple droplets affects
scattering only at long wavelengths. If the temperature is not
small compared to the melting temperature, the droplet phase will be a
liquid and interference from scattering off different
droplets are small for neutrino energies $E_\nu \gsim
(1/R_W)$. However, multiple droplet scattering cannot be neglected for
$E_\nu \lsim 1/R_W$. The effects of other droplets is to cancel
scattering in the forward direction, because the interference is
destructive except at exactly zero degrees, where it produces a change
in the index of refraction of the medium. These effects are usually
incorporated by multiplying the differential cross section
equation (\ref{diff}) by the static form factor of the medium
\begin{equation}
S(q)= 1 + N_D \int d^3r \exp{i{\bf q}.{\bf r}}~(g(r)-1) \,,
\end{equation}
where $g(r)$ is the radial distribution function of the droplets.  The
droplet correlations, which determine $g(r)$, are measured in terms of
the dimensionless Coulomb number $\Gamma=Z^2e^2/(8\pi R_W kT)$. Due to
the long-range character of the Coulomb force, the role of screening
and the finite droplet size, $g(r)$ cannot be computed
analytically. We use a simple ansatz for the radial distribution
function $g(r< R_W) = 0$ and $g(r>R_W)=1$. For this choice, $S(q)$ is
independent of $\Gamma$. Monte Carlo calculations \cite{CJH} of a
simple one component plasma indicate that this choice of $S(q)$ is
adequate for neutrino energies of interest.

The simple ansatz for $g(r)$ is equivalent to subtracting, from the weak
charge density $\rho_W$, a uniform density which has the same total weak
charge $N_W$ as the matter in the Wigner-Seitz cell. Thus, effects due
to $S(q)$ may be incorporated by replacing the form factor $F(q)$ by
\begin{eqnarray}
F(q) \rightarrow \tilde{F}(q) = F(q) - 3~
\frac{\sin{qR_{W}} - (qR_{W})\cos{qR_{W}}}{(q R_{W})^3}  \,.
\label{formc}
\end{eqnarray}
The neutrino--droplet differential cross section per unit volume
is then 
\begin{equation}
\frac{1}{V}\frac{d\sigma}{d\cos{\theta}}=
N_D~\frac{E_\nu^2}{16\pi} G_F^2 N^2_W(1+\cos{\theta}) \tilde{F}^2(q) \,.
\label{diff1}
\end{equation}
Note that even for small $N_D$, the factor $N_W^2$, which in the case
of nuclei is proportional to $N$, serves to enhance the droplet
scattering.

This concludes our overview of some of the technical tools that are
needed and available to address media-related issues.

\subsection{Neutrino Mean Free Paths: Examples}

The differential cross section in equation (\ref{formalneutral}) is
required in multi-energy group neutrino transport calculations (see
Section 5.1). However, more approximate neutrino transport schemes
(such as those described in Section 5.2.2) utilize the total cross section
per unit volume (or equivalently the inverse mean free path),
integrated over the angle $\theta$ and energy transfer $\omega$ in
equation (\ref{formalneutral}), as a function of the neutrino energy.
Examples of such neutrino scattering (that is common
to all neutrino species) and absorption mean free paths for conditions
relevant to the deleptonization and cooling epochs of PNSs
are considered below.  The equations of state used in these
calculations are described in \cite{prak97a}.

\begin{figure}
\begin{center}
\includegraphics[scale=.6]{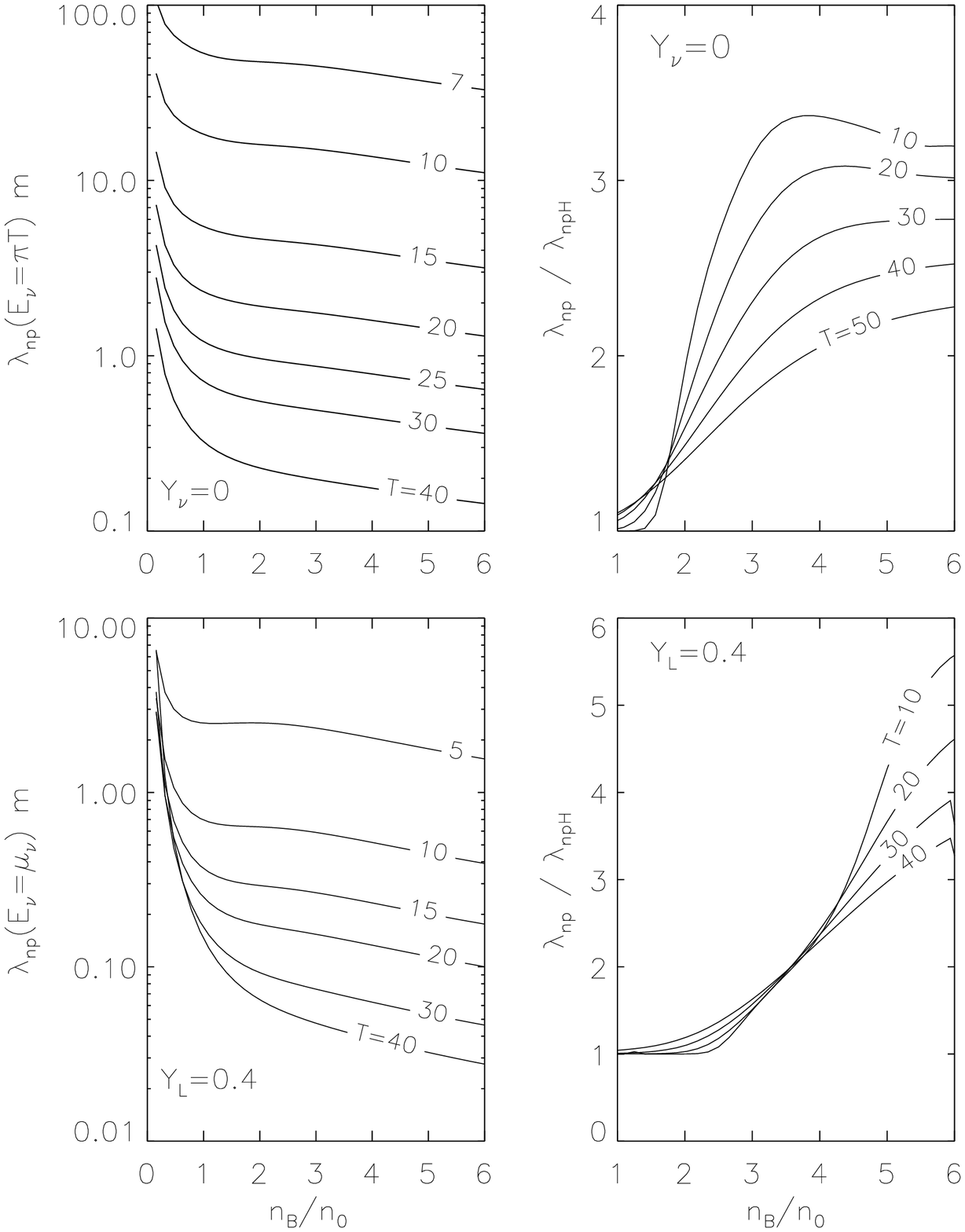}
\end{center}
\caption{Neutrino mean free paths in matter with nucleons only (left
panels).  Right panels show ratios of mean free paths in matter
without and with hyperons.  Top panels show scattering mean free paths
(common to all neutrino species).  Bottom panels show $\nu_e$ mean
free paths including absorption reactions.  The figure is from
Ref. \cite{redd98}.}
\label{sig}
\end{figure}

\paragraph{A. Effects of Composition:}
Under degenerate conditions even modest changes to the composition
significantly alter the neutrino mean free paths.  In Figure~\ref{sig}
the top (bottom) panels show the scattering (scattering plus
absorption) mean free paths in neutrino-free (neutrino-trapped)
matter.  The left (right) panels show results for selected
temperatures (for the neutrino energy $E_{\nu}=\pi T$) in matter
without (with) hyperons.  The presence of hyperons significantly
decreases the mean free paths in both environments because of the 
additional available channels. 

\begin{figure}
\begin{center}
\leavevmode
\includegraphics[scale=.35]{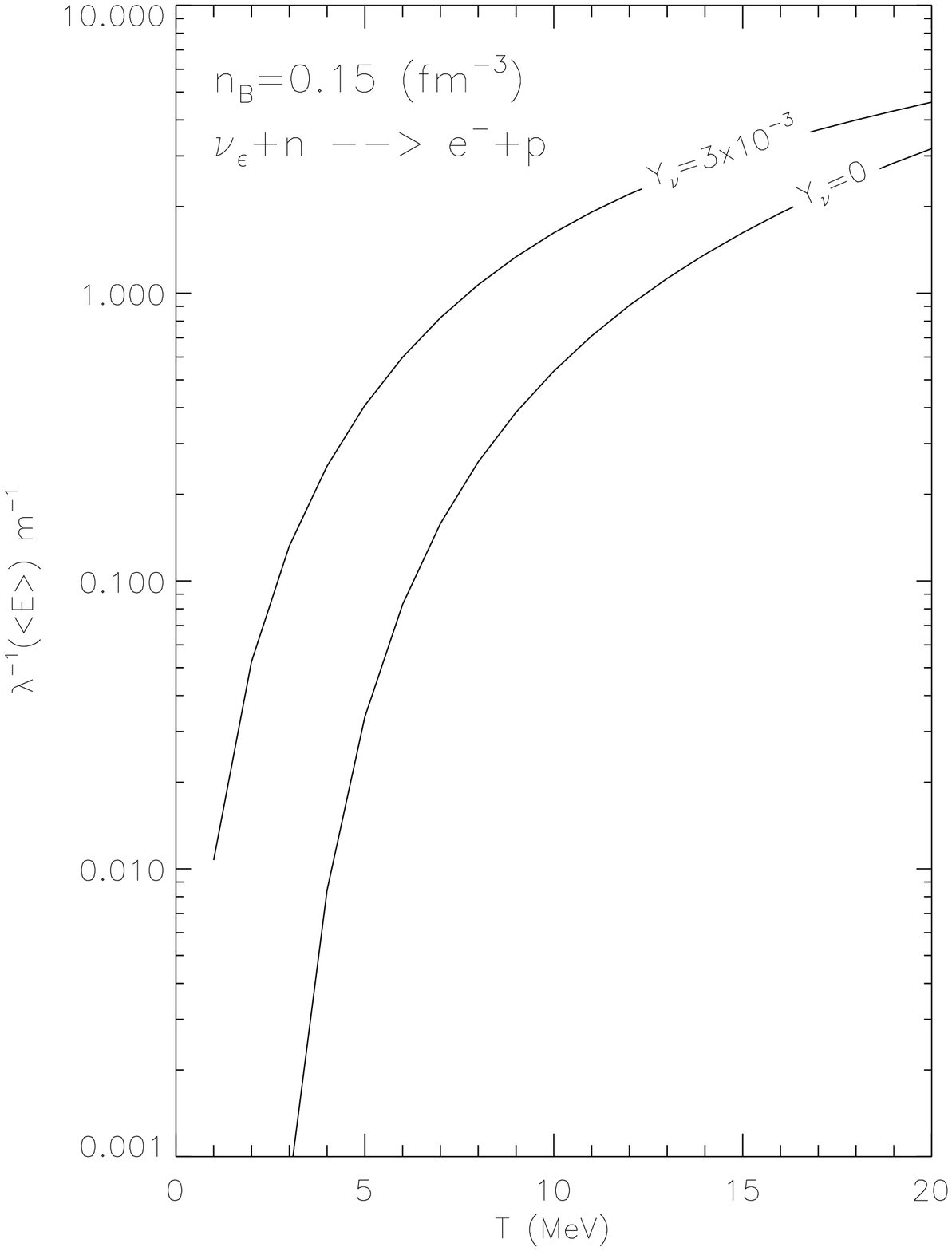}
\includegraphics[scale=.35]{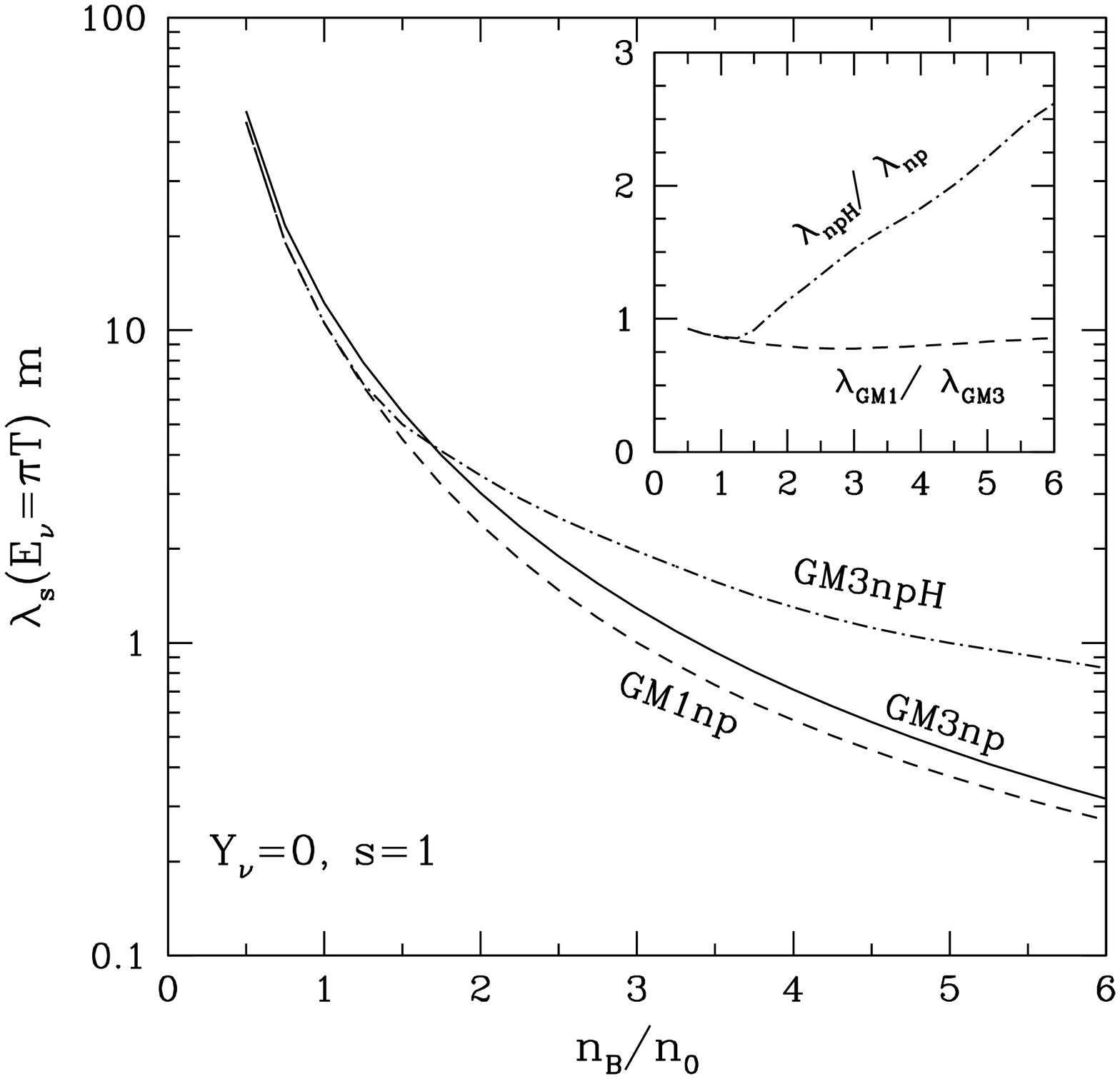}
\end{center}
\caption{ Left: Charged current inverse neutrino mean free
paths. Right: Comparison of scattering mean free paths in neutrino
poor matter at fixed entropy in matter containing
nucleons and also hyperons.  The figure is from Ref. \cite{redd98}.}
\label{csig}
\end{figure}

During the deleptonization stage, charged current reactions in fact
dominate scattering reactions.  At zero temperature, reactions like
the direct Urca process $\nu + n \leftrightarrow e + p$ depend
sensitively on the proton fraction $Y_p$~\cite{LPPH91}.  Kinematic
restrictions require $Y_p$ to be larger than $11-14\%$ (this is called
the direct Urca threshold).  At early times, when large numbers of
trapped neutrinos are present, these reactions proceed without
hindrance.  After several tens of seconds, however, $Y_p$, which
depends sensitively on the density dependence of the nuclear symmetry
energy, decreases to its cold, catalyzed value.  In field-theoretical
models, the symmetry energy is largely controled by the $\rho$-meson
exchange which increases strongly with density, and establishes a
typical critical density $n_c=2\sim3 n_0$. However, at finite
temperature, the equilbrium $Y_p$ and average $E_\nu$ values are
larger than at zero temperature, enabling the charged current
reactions to proceed even below $n_c$.  Figure~\ref{csig} shows that
this is indeed the case even at relatively low temperatures ($T\sim
3-5$) MeV for a baryon density $n_B=0.15~{\rm fm}^{-3}$. Thus, Urca
processes dominate the opacity until very late times.

\paragraph{B. Effects of Strong and Electromagnetic Correlations:}

\begin{figure}
\begin{center}
\includegraphics[scale=0.75]{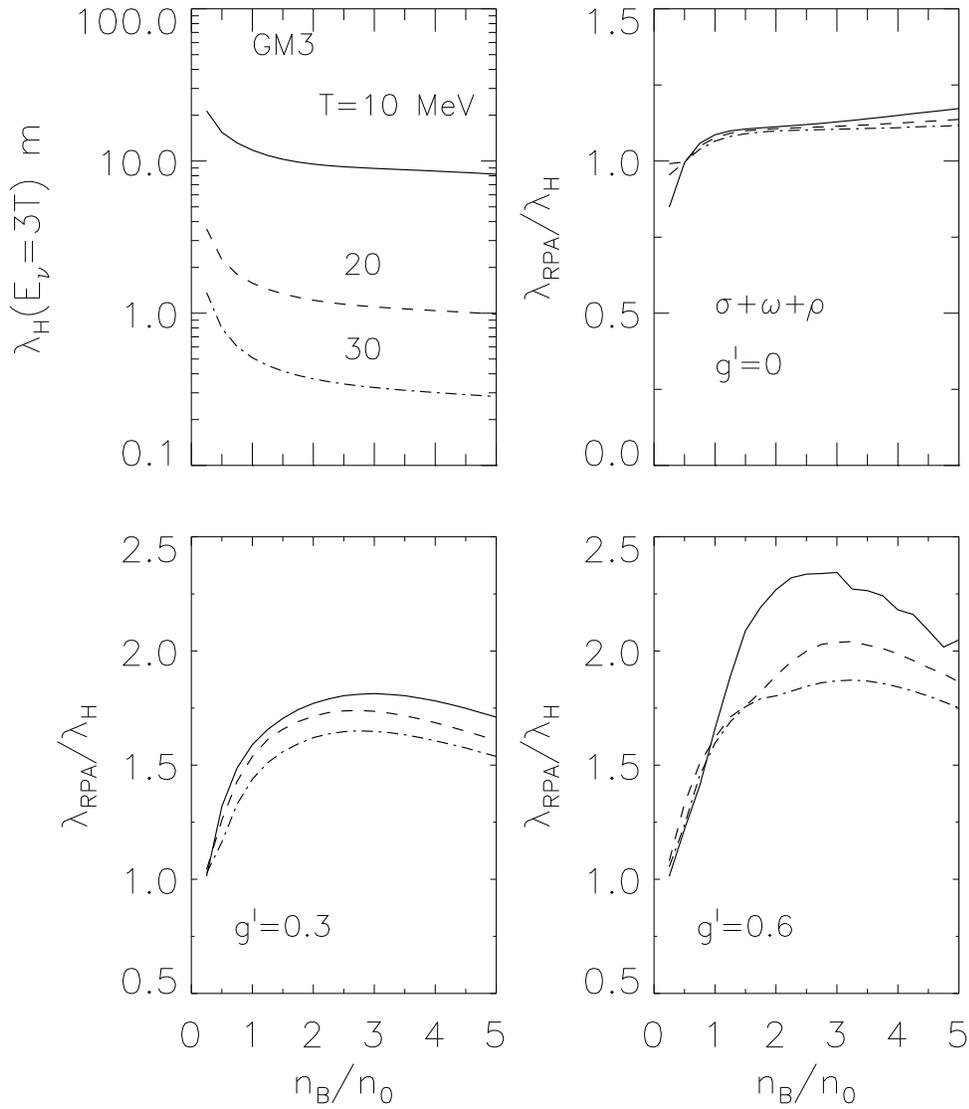}
\caption[]{ The density and temperature dependences of the neutral current mean
free paths for neutrino free matter in the field-theoretical
model GM3. The upper left panel shows the Hartree results for the case
$E_\nu=3T$.  The influence of the spin correlations introduced via the Migdal
parameter $g^\prime$ is strong, as can be deduced from the results shown in
the upper right and bottom panels.  The figure is from
Ref. \cite{redd99a}.}
\label{rlam}
\end{center}
\end{figure}

The RPA and Hartree scattering mean free paths of thermal neutrinos in
neutrino-free matter are compared in Figure~\ref{rlam}.  Correlations
are generally more important with increasing density and decreasing
temperature. The density dependence is chiefly due to variations in
the effective baryon mass, which is controlled by the Fermi Liquid
parameter $g^\prime$.  The RPA correlations, even for a large $g'$,
will typically increase the mean free path by a factor $2-2.5$.

\begin{figure}
\begin{center}
\includegraphics[scale=0.65]{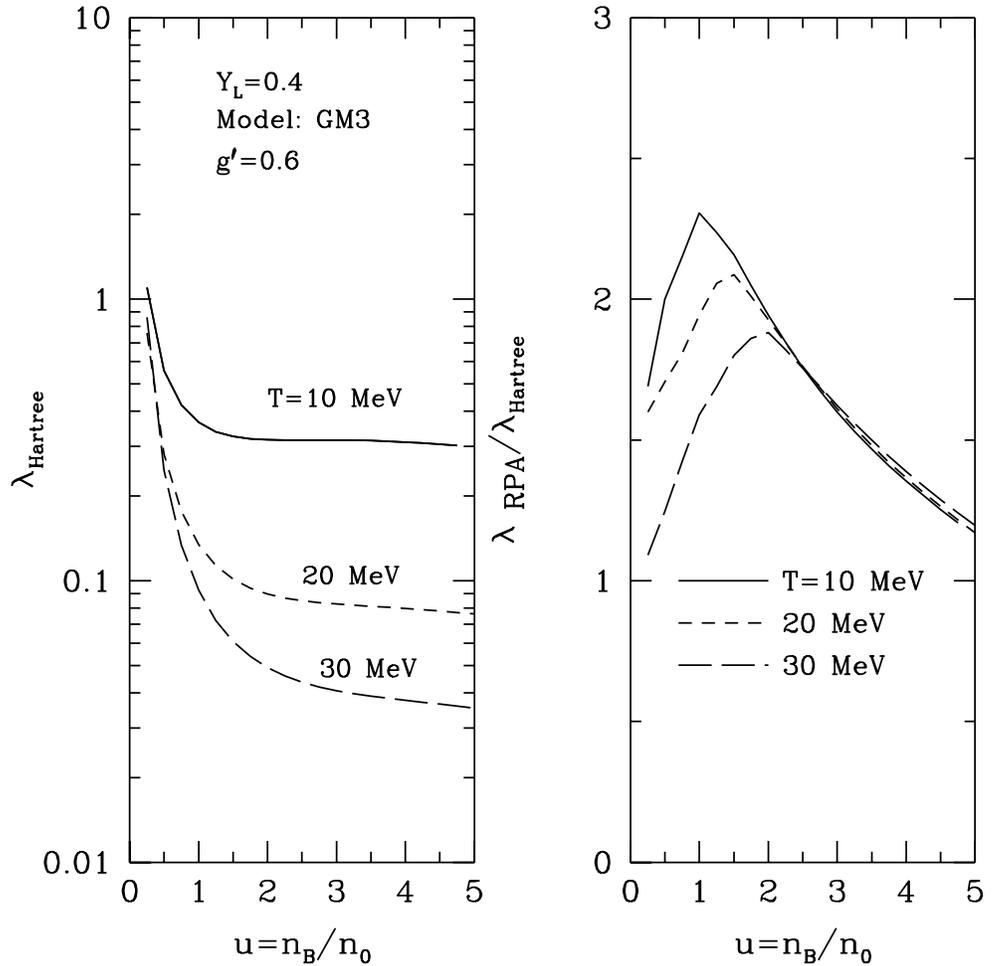}
\caption{The density and temperature dependences of the charged current
neutrino  mean free path in $\beta-$stable matter for the GM3 model
assuming $Y_L=0.4$. Results for the Hartree approximation (left panel) are
compared with those including RPA corrections (right panel) with
$g^\prime=0.6$.  The figure is from
Ref. \cite{redd99a}.}
\label{arsig_y4}
\end{center}
\end{figure}

The RPA and Hartree absorption mean free paths are compared in Figure
\ref{arsig_y4} for neutrino-trapped matter.  The qualitative trends
for neutrino-free matter are similar, although the largest
enhancements due to correlations are a factor of 5 in the
neutrino-free case instead of about 2 in the neutrino-trapped case.
In both cases, the results are quite similar to those for the
scattering case shown in Figure~\ref{rlam}.

\paragraph{C. Mean Free Paths in Quark Matter:}

\begin{figure}
\begin{center}
\leavevmode
\includegraphics[scale=0.7]{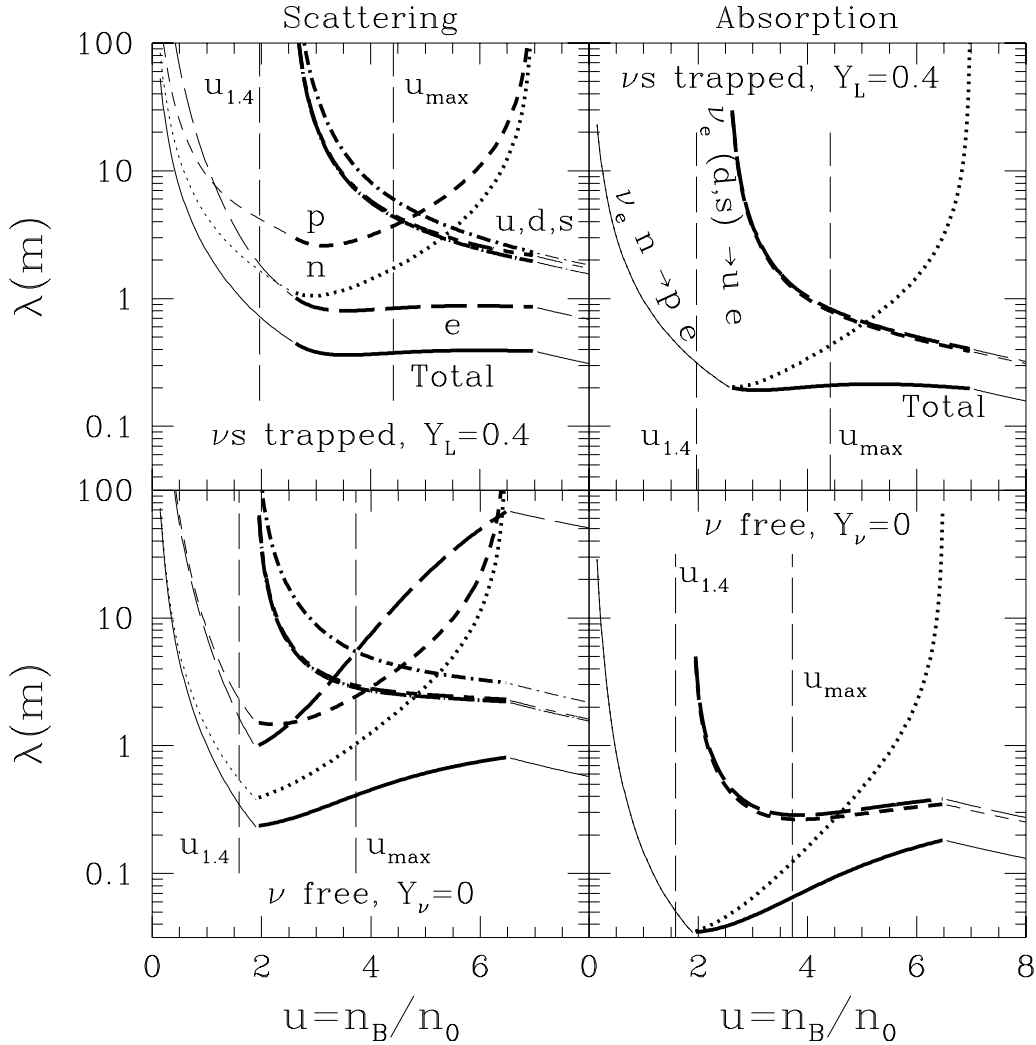}
\caption{$\nu_{e}$ mean free paths from various particles in matter
containing a mixed phase of quarks and hadrons.  Thick lines show the
extent of the mixed phase region.  Left (right) panels show scattering
(absorption) mean free paths.  The upper (lower) panels correspond to
the neutrino-trapped (neutrino-free) era. Vertical dashed lines
labelled $u_{1.4}$ and $u_{max}$ indicate the central densities of 1.4
M$_{\odot}$ and maximum mass (2.22 M$_{\odot}$ for the upper panels
and 1.89 M$_{\odot}$ for the lower panels) stars, respectively.}
\label{cs3l}
\end{center}
\end{figure}

An intriguing possibility is that sufficiently dense matter contains
deconfined quarks.  It is likely that quark matter, if it exists, will
comprise one component of a mixed phase with hadrons~\cite{Glen}.
Within the mixed phase, the thermodynamic and chemical potentials will
be rather different than in ordinary hadronic matter of the same
density.  As a result, significant effects on neutrino opacities are
anticipated.  Figure~\ref{cs3l} shows scattering and absorption mean
free paths, for the cases of matter with trapped neutrinos and
neutrino-free matter, in which a mixed phase occurs~\cite{SPL01}.  The
vertical dashed lines indicate the central densities of 1.4 M$_\odot$
stars and maximum mass stars, respectively; densities above the
right-most vertical line cannot be achieved in any neutron star.  The
thin curves show the mean free paths for the respective matter
(hadrons or quarks) in their pure phases, while their values within
the mixed phase are indicated by thick curves.  The chief consequence
of a mixed phase is that it reverses the trend of pure hadronic matter
to have steadily decreasing mean free paths with increasing densities.
In other words, matter within the mixed phase region is substantially
more transparent than matter without such a transition, whether or not
the matter is composed of pure hadrons or quarks.  Partly, this result
is a consequence of the fact that the entropy is held fixed in these
panels: quark matter has a higher specific heat than hadronic matter
and favors a lower temperature.  In the presence of a mixed phase, the
temperature is smaller than in its absence, and the neutrino cross
sections, which increase roughly as $T^2$, decrease.

\paragraph{D. Droplet Scattering In Heterogeneous Phases:}

\begin{figure}
\begin{center}
\leavevmode
\includegraphics[scale=0.34,angle=0]{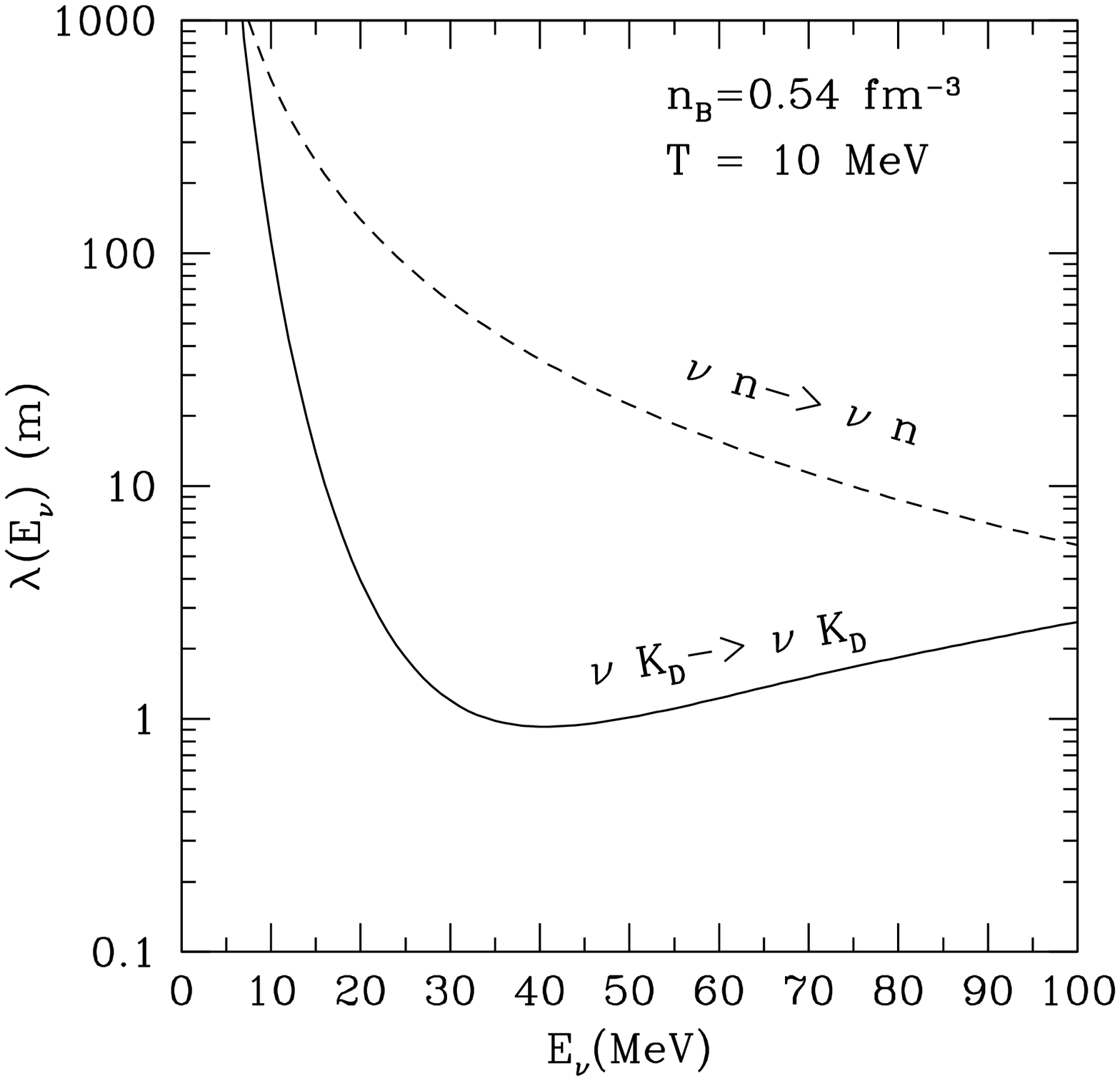}
\includegraphics[scale=0.32,angle=0]{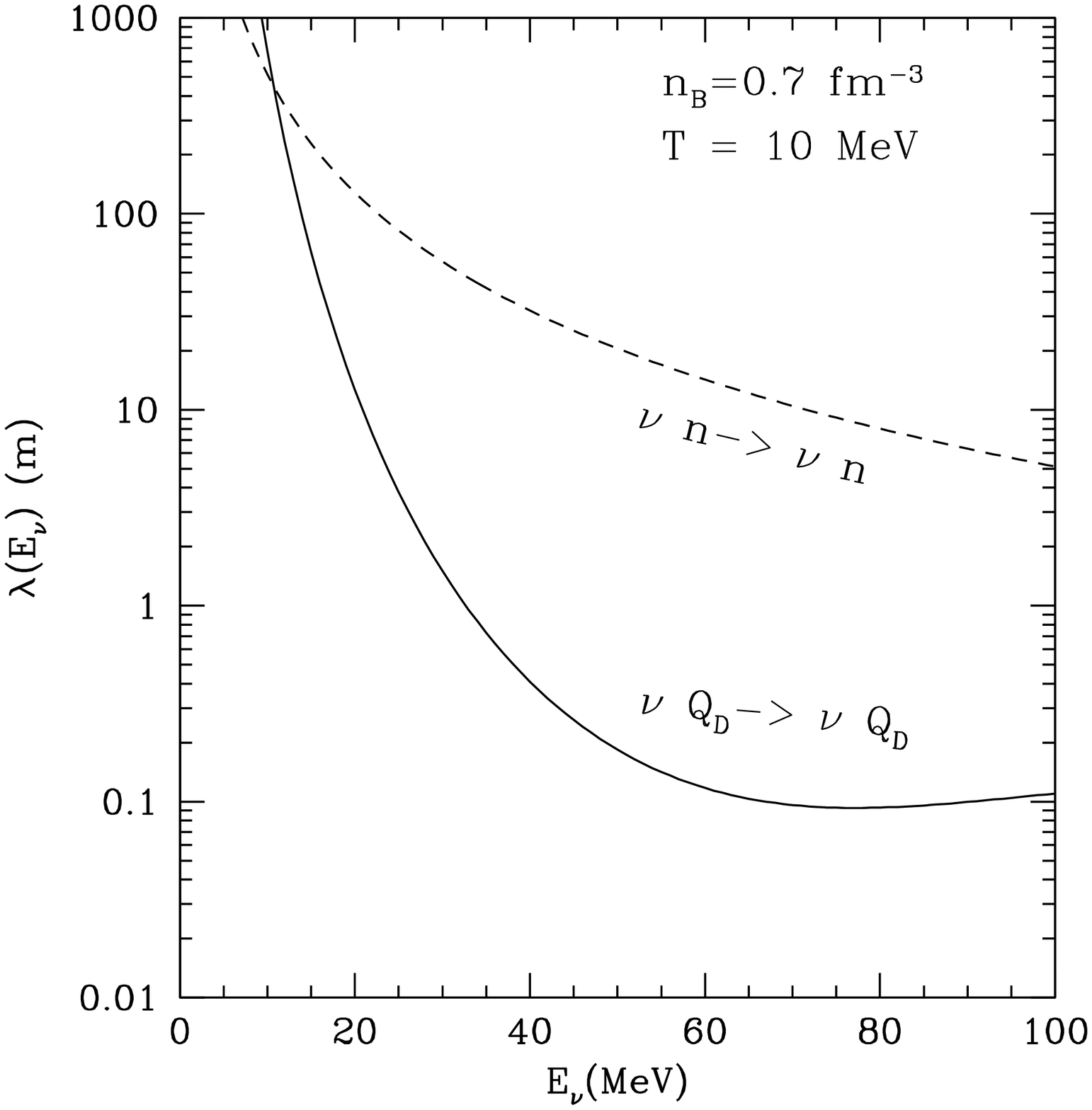}
\caption{Transport neutrino mean free paths in heterogeneous media.
Solid lines are for matter in a mixed phase containing kaons (left
panel) and quarks (right panel), and dashed curves are for uniform
matter.  This figure is from Ref. \cite{RBP00}.}
\label{mpaths}
\end{center}
\end{figure}

Models of first order phase transitions in dense matter \cite{drops}
provide the weak charge and form factors of the droplets and permit
the evaluation of $\nu$--droplet scattering contributions to the
opacity of the mixed phase \cite{RBP00}. For first order kaon
condensate and quark-hadron phase transitions, the neutrino mean free
paths in the mixed phase are shown in the left and right panels of
Figure \ref{mpaths}, respectively.  The transport cross sections in
diffusive transport are usually differential cross sections weighted
by the angular factor $(1-\cos{\theta})$.  The kaon droplets are
characterized by radii $R_N\sim7$ fm and inter-droplet spacings
$R_W\sim20$ fm, and enclose a net weak vector charge $N_W \sim
700$. The quark droplets are characterized by $R_N\sim5$ fm and
$R_W\sim11$ fm, and an enclosed weak charge $N_W \sim 850$.  For
comparison, the neutrino mean free paths in uniform neutron matter at
the same density and temperature are also shown.  A large
magnification in the opacity exists for $E_\nu \sim \pi T$.  At much
lower energies, the inter-droplet correlations tend to screen the weak
charge of the droplet, and at higher energies the coherence is
attenuated by the droplet form factor.  Similar effects occur in the
case of coherent scattering from nuclei \cite{BURLAT}.

\section{NEUTRINOS IN THE EARLY UNIVERSE}

\subsection{Big Bang Nucleosynthesis}

An interesting application of neutrino transport in a dense medium
lies in the study of neutrino oscillations in that phase of the early
universe immediately prior to and during Big Bang Nucleosynthesis
(BBN) \cite{c1}. Neutrinos play two very important roles during BBN.

First, light neutrinos will be relativistic during the nucleosynthetic 
epoch, and so contribute
to driving the expansion of the universe. Under radiation dominance,
the expansion rate or Hubble parameter is given by
\begin{equation}
{\cal H} = \left( \frac{4\pi^3}{45} g_{*} \right)^{1/2} \frac{T^2}{M_P}\,,
\label{H}
\end{equation}
where natural units $\hbar = c = 1$ have been employed, $T$ is temperature
and $M_P$ is the Planck
mass. The effective number of massless degrees of freedom $g_{*}$ is
\begin{equation}
g_{*} = \sum_{{\rm bosons}} g_i \left( \frac{T_i}{T} \right)^4 +
\frac{7}{8} \sum_{{\rm fermions}} g_i \left( \frac{T_i}{T} \right)^4\,,
\label{g}
\end{equation}
where $g_i$ counts the internal states of species $i$. Light neutrinos
contribute to the sum over fermionic species. If particle of type $i$
is in thermal contact with the plasma, then $T_i = T$. If the
expansion rate is smaller than the rate for a given process (e.g.\ a
nuclear reaction), then that process will be dynamically important.

The second BBN role arises from electron neutrinos and
antineutrinos transforming protons into neutrons and vice-versa through the
reactions
\begin{equation}
\nu_e n \leftrightarrow e^{-} p,\qquad \overline{\nu}_e p \leftrightarrow
e^{+} n\,.
\label{npreactions}
\end{equation}
When chemical equilibrium obtains, the ratio of neutron to proton
number densities is given by
\begin{equation}
\frac{n_n}{n_p} = 
\exp\left(-\frac{m_n - m_p}{T} - \frac{\mu_{\nu_e}}{T}\right)\,,
\end{equation}
where $\mu_{\nu_e}$ is the electron neutrino chemical potential. A
crucial event just prior to BBN is {\it weak freeze out}, when the
reaction rates for equation\ (\ref{npreactions}) drop below ${\cal
H}$. These reactions are then no longer rapid enough to maintain the
$n/p$ ratio at its equilibrium value, and neutron decay becomes
important. The $n/p$ ratio determines the amount of primordial $^4$He
synthesized, because to a good first approximation all available
neutrons are incorporated into this species.  ``Standard BBN'' sees
$\mu_{\nu_e}$ arbitrarily set to zero.  However, there is no known
fundamental reason to require $\mu_{\nu_e} = 0$.  Neutrino
oscillations can lead to time dependent neutrino chemical potentials,
with a direct impact on BBN through the $n/p$ ratio.

The mean total collision rate $\langle \Gamma_{\alpha} \rangle$ for
neutrino $\nu_{\alpha}$ ($\alpha = e,\mu,\tau$)
in the epoch of the early universe after $\mu/\overline{\mu}$ disappearence is
\begin{equation}
\langle \Gamma_{\nu_\alpha} \rangle \simeq \kappa_{\alpha} G_F^2 T^5\,,
\label{meanweakrates}
\end{equation} 
where $\kappa_e \simeq 4$, $\kappa_{\mu,\tau} \simeq 2.9$ and $G_F$ is the
Fermi constant. The average is taken
over a Fermi-Dirac (FD) distribution
\begin{equation}
N^{eq}(p, \mu_{\nu_\alpha}, T) \equiv \frac{p^2}{2\pi^2} 
f^{eq}(\frac{p}{T}, \frac{\mu_{\nu_\alpha}}{T}) 
= \frac{1}{2\pi^2} \frac{p^2}{e^{\frac{p - \mu_{\nu_\alpha}}{T}} + 1}\,,
\label{FD}
\end{equation}
where $p \simeq E$ is the magnitude of the three-momentum of the
neutrino.
The chemical potential $\mu_{\nu_\alpha}$ has been
arbitrarily equated to zero in computing 
the righthand side of equation (\ref{meanweakrates}). If
antineutrinos are zero chemical potential FD distributed, then
their mean collision rates are also given by
equation (\ref{meanweakrates}). Note that the total 
equilibrium number density $n_{\nu_\alpha}$ is given by
\begin{equation}
n_{\nu_\alpha}(\mu_{\nu_\alpha},T) = \int_0^{\infty} 
N^{eq}(p,\mu_{\nu_\alpha},T)\, dp\,.
\end{equation}

If we take the constituents of the plasma at this epoch to be free of
exotic states such as light sterile neutrinos, light neutral spin-$0$
bosons and the like, then from equations (\ref{H}) and (\ref{g})
the expansion rate of the universe is
\begin{equation}
{\cal H} \simeq 5.44 \frac{T^2}{M_P}\,,
\label{exprate}
\end{equation}
with $g_{*} = 10.75$ under the stated assumptions.

Neutrino collisions will be very
important for all temperatures above the decoupling temperature,
$T_{\nu_\alpha}$, defined through
\begin{equation}
\kappa_{\alpha} G_F^2 T_{\nu_\alpha}^5 \simeq 5.44 
\frac{T^2_{\nu_\alpha}}{M_P} \Rightarrow
T_{\nu_\alpha} \simeq  1\ {\rm MeV}\,.
\end{equation}
Above about 1 MeV, collisions keep neutrinos in thermal equilibrium with
the electrons, positrons, photons and the other neutrinos and
antineutrinos. Also, neutrino oscillations will be collision affected in
this regime, necessitating the use of the Quantum Kinetic Equation
formalism \cite{MandT,raff2,c2,c3,c4,c5,c6,c7}. 
Since BBN begins at about $0.4$ MeV, the neutrinos
are expected
to be propagating in a collisionless way during the BBN epoch
itself. However, the
initial conditions for BBN (such as the $n/p$ ratio at freeze-out) depend
on the collision-affected neutrino dynamics of the previous epoch. 

Naively, one would hope to focus on oscillation dynamics involving the
three known neutrinos $\nu_{e,\mu,\tau}$. However, most of the
interesting work has focussed on active-sterile neutrino oscillations.
The reasons for this are, first, that sterile neutrinos can have
dramatic consequences for BBN, and, second, the Quantum Kinetic
Equations for an active-sterile system, equation (\ref{4}), are much
simpler than for its active-active counterpart, equation (\ref{5}).

Active-active oscillations are often neglected in the literature
because of the common assumption that all neutrino chemical potentials
are zero, or at least extremely small. If the chemical potentials are
in fact nonzero, then an important consequence is the inequality of
neutrino and antineutrino number densities of a given flavor. Define
the {\it $\alpha$-flavor neutrino asymmetry} $L_{\nu_{\alpha}}$ by
\begin{equation}
L_{\nu_{\alpha}}  \equiv  
\frac{n_{\nu_{\alpha}} - n_{\overline{\nu}_{\alpha}}}{n_{\gamma}}\,.
\end{equation}
When thermal equilibrium holds this evaluates to
\begin{equation}
L_{\nu_{\alpha}} \simeq  \frac{1}{24\zeta(3)}\left[
\pi^2(\xi_{\nu_\alpha} - \overline{\xi}_{\nu_\alpha}) - 
6 (\xi_{\nu_\alpha}^2 - \overline{\xi}_{\nu_\alpha}^2)\ln 2 +
(\xi_{\nu_\alpha}^3 - \overline{\xi}_{\nu_\alpha}^3) \right]\,,
\label{Ltomurelation}
\end{equation}
where
$\xi_{\nu_\alpha} \equiv \mu_{\nu_\alpha}/T$, 
$\overline{\xi}_{\nu_\alpha} \equiv
\mu_{\overline{\nu}_\alpha}/T$ are dimensionless chemical potentials. This
equation is exact if 
$\overline{\xi}_{\nu_\alpha} = - \xi_{\nu_\alpha}$, otherwise
it is a good approximation for $|\xi| \stackrel{<}{\sim} 1$. It is
convenient to scale the neutrino number density with the photon number
density, and to consider the $\xi$'s in place of the $\mu$'s, because the
scaled quantities do not red-shift. While processes such as $\nu_{\alpha}
\overline{\nu}_{\alpha} \leftrightarrow e^{+} e^{-} \leftrightarrow \gamma
\gamma$ are occurring quickly relative to the expansion rate, the
neutrinos and antineutrinos are in chemical equilibrium which requires
$\xi_{\nu_\alpha} + \overline{\xi}_{\nu_\alpha} = \xi_{e^{-}} + \xi_{e{+}} =
0$, so equation (\ref{Ltomurelation}) is then exact. 

The assumption that $\nu/\overline{\nu}$ chemical potentials are
zero, or perhaps of the tiny magnitude motivated by the baryon/electron
asymmetries, is one of the defining features of standard BBN. 
However, the equal number densities so implied for all active
neutrino species render neutrino oscillations
cosmologically uninteresting, simply
because equal distributions would just be exchanged through the
oscillations. We will therefore concentrate on cases featuring unequal
distribution functions. A few well motivated situations of this sort are:
\begin{enumerate}
\item {\it Sterile neutrinos.}  They would decouple very early, with
subsequent reheating processes greatly diluting the putative sterile
component of the plasma, so that by the epoch immediately prior to BBN
their number density would be effectively zero. Active-sterile
oscillations could then repopulate the plasma with sterile neutrinos.
\item {\it Mirror neutrinos.}  They are similar to sterile neutrinos,
except that they have weak-scale self-interactions, and their small
initial number densities are justified differently
\cite{c8,c9,c10,c11}.
\item { \it Active neutrinos in the presence of large chemical
potential differences between the flavors.} As we will see,
active-sterile oscillations can create reasonably large chemical
potentials during the epoch preceeding BBN \cite{c12}. After this has
happened, other oscillation modes, including active-active channels,
can reprocess the flavor of the lepton asymmetry \cite{c13}. In
addition, large chemical potentials can be created during much earlier
epochs, for example by an Affleck-Dine mechanism
\cite{affleckdine,gelmini}, which can be similarly reprocessed.
\end{enumerate}

\subsection{Active-Sterile Oscillations: Formalism}

\subsubsection{Quantum Kinetic Equations}

Equations (\ref{5}) and (\ref{6}) 
describe the evolution of the 1-body reduced density matrix $\rho$
for an active-sterile system. The antineutrino system is described by a similar 
equation for $\overline{\rho}$. 

The diagonal entries of $\rho$ ($\overline{\rho}$) are appropriately
normalized distribution functions for $\nu_{\alpha}$
($\overline{\nu}_{\alpha}$) and $\nu_s$ ($\overline{\nu}_s$): 
\begin{eqnarray}
N_{\nu_\alpha}(p,T) & = & \frac{1}{2} [P_0(y,T) + P_z(y,T)] ~N^{eq}(p,0,T)\,,
\ \\ N_{\nu_s}(p,T) & = & \frac{1}{2} [P_0(y,T) - P_z(y,T)] ~N^{eq}(p,0,T)\,,
\label{relnums}
\end{eqnarray}
where $N^{eq}(p,0,T)$ is the often-used reference distribution function.
The antineutrino distribution functions are given similarly. Since we have
defined $P_{0,z}$ to be ratios of distribution functions, there is no
term in the QKEs related to the expansion of the universe. It is
understood, however, that a neutrino which has momentum $p$ at temperature
$T$ will red-shift to momentum $p'$ at temperature $T'$ such that
$p/T = p'/T'$. In other words, the scaled momentum
\begin{equation}
y \equiv \frac{p}{T}
\end{equation}
is the natural and most convenient variable.
Note that $y$ has nothing to do with the unit vector $\hat{y}$.

The function ${\bf V}(y,T)$, describing the quantally coherent part of the
evolution, is given by
\begin{equation}
{\bf V}(y,T) = \beta(y,T)\hat{x} + \lambda(y,T)\hat{z}\,,
\end{equation}
with
\begin{equation}
\beta(y,T)  =  \frac{\Delta m^2}{2yT} \sin2\theta_0,\quad
\lambda(y,T)  =  - \frac{\Delta m^2}{2yT}\cos2\theta_0 +
V_{\alpha}(y,T)\,,
\end{equation}
where
$\Delta m^2$ and $\theta_0$ are, respectively, the mass-squared 
difference and vacuum mixing angle
for $\nu_{\alpha} - \nu_s$ oscillations.
The mass eigenstate neutrinos $\nu_{a,b}$ are defined by
$\nu_{\alpha} = \cos\theta_0\nu_a + \sin\theta_0\nu_b$,
$\nu_s = -\sin\theta_0\nu_a + \cos\theta_0\nu_b$,
with $\theta_0$ defined so that $\cos2\theta_0 \ge 0$ and 
$\Delta m^2 \equiv m_b^2 -
m_a^2$. The function $V_{\alpha}$ is the effective 
matter potential \cite{c14,c15}.
Calculated to the appropriate order for our applications, it is given by
\cite{c16}
\begin{equation}
V_{\alpha}(y,T) = \frac{\Delta m^2}{2yT}[- a(y,T) + b(y,T)]\,,
\label{Valpha}
\end{equation}
with the dimensionless functions $a(y,T)$ and $b(y,T)$ being
\begin{equation}
a(y,T)  =  - \frac{4\zeta(3)\sqrt{2}}{\pi^2}\frac{G_F T^4 y}{\Delta m^2}
L^{(\alpha)}\,,\quad
b(y,T)  =  - \frac{4\zeta(3)\sqrt{2} A_{\alpha}}{\pi^2} \frac{G_F T^6
y^2}{\Delta m^2 m_W^2}\,,
\end{equation}
where $m_W$ is the $W$-boson
mass, $A_e \simeq 17$, $A_{\mu,\tau} \simeq 4.9$  and the $\alpha$-type
{\it effective neutrino asymmetry} is defined by
\begin{equation}
L^{(\alpha)} = L_{\nu_\alpha} + L_{\nu_e} + L_{\nu_\mu} + L_{\nu_\tau} + \eta.
\label{L}
\end{equation}
Observe that ${\bf V}$ depends on $\rho$ through
the dependence of $a$ on $L_{\nu_\alpha}$, making
the asymmetry evolution non-linear. The
quantity $\eta$ is a small term related to the cosmological
baryon-antibaryon asymmetry. For antineutrinos, the corresponding function
$\overline{{\bf V}}$ is obtained from ${\bf V}$ by replacing $L^{(\alpha)}$
by $-L^{(\alpha)}$.

The Mikheyev-Smirnov-Wolfenstein (MSW)
resonance conditions are \cite{c14,c15,c16} 
\begin{equation}
\cos 2\theta_0 + a(y,T) - b(y,T)  = 0,\quad \cos 2\theta_0 - a(y,T) -
b(y,T) = 0\,,
\label{res}
\end{equation}
for neutrinos and antineutrinos, respectively.  It is important to
appreciate that the resonance conditions at a given temperature are
met only for neutrinos (antineutrinos) of a certain momentum $y_{res}$
$(\overline{y}_{res})$.

The decoherence function is $D(y,T) = \Gamma_{\nu_\alpha}(y,T)/2$,
with $\Gamma_{\nu_\alpha}(y,T)$ being the total collision rate for
$\nu_{\alpha}$'s with
momentum $y$. When thermal and chemical equilibria
hold,
\begin{equation}
\Gamma_{\alpha}(y,T) = \kappa_{\alpha} \frac{180 \zeta(3)}{7 \pi^4} G_F^2
T^5 y + O(L) \,,
\label{Gammap}
\end{equation}
where the $O(L)$ term takes care of possible asymmetries in the medium.
The antineutrino decoherence function is obtained by reversing the signs of
all the asymmetries.

The function $D$ is in general a dynamical quantity because the collision
rates depend on the neutrino
distribution functions and hence on $\rho$. However, it
approximates to the external function displayed above provided that
thermal equilibrium obtains and all lepton numbers are small (but not
necessarily tiny). This greatly simplifies the numerical solution of the
equations.  Fermi factors have been neglected in the calculation of $D$,
and also the repopulation function $R$ discussed below. 
  
The repopulation function $R$ is approximately
given by
\begin{equation}
R(y,T) = \Gamma_{\alpha}(y,T)\left\{
\frac{N^{eq}(p,\xi_{\alpha},T)}{N^{eq}(p,0,T)}
- \frac{1}{2} [P_0(y,T) + P_z(y,T)] \right\}\,,
\label{explicitR}
\end{equation}
when all distribution functions except that for $\nu_{\alpha}$ are of
FD form, and the $\nu_{\alpha}$ distribution is approximately FD.  The
physical interpretation of this expression is that all weak
interaction processes involving $\nu_{\alpha}$ are tending to send its
actual distribution function towards equilibrium FD form. For
antineutrinos, $\overline{R}$ is obtained from $R$ by replacing
$\xi_{\nu_\alpha}$ by $\overline{\xi}_{\nu_\alpha}$ and the $P$'s by
$\overline{P}$'s.

\subsubsection{The Adiabatic Limit}

The Quantum Kinetic Equations in principle provide a complete solution
to oscillating neutrino dynamics in this context. Indeed, numerical
solutions to these equations have been computed for certain situations
\cite{c11,c17,c18,c19}.  (Numerical work which considers the
simplified case where the full energy spectrum of neutrinos is
replaced by the mean momentum can be found, for instance, in Refs.\
\cite{c7,c12,c20,c21,c22,buras}. For calculations in a collisionless
regime, see Refs.\ \cite{c23prime,c23,c24}.)  Before reviewing some of
these results, it will be helpful to extract some analytical
understanding of important features of the dynamics
\cite{c12,c17,deviantearly,c25,c26}.  It turns out that taking the
adiabatic limit is a useful analytical strategy.  The functions
$\beta$, $\lambda$, $D$ and $R$ all depend on time or temperature. The
adiabatic limit is useful when the rates of change of these quantities
are sufficiently small \cite{deviantearly}.

Consider first the higher temperature regime where collisions maintain the
$\nu_{\alpha}$'s (and $\overline{\nu}_{\alpha}$'s) close to thermal
equilibrium, so that we can adopt
the {\it instantaneous repopulation approximation}, whereby
\begin{equation}
\frac{P_0(y,T) + P_z(y,T)}{2} \simeq 
\frac{N^{eq}(p,\xi_{\alpha},T)}{N^{eq}(p,0,T)}
\label{instrepop}
\end{equation}
is maintained at all times, though possibly with a time-dependent chemical
potential. It turns out that the further approximation $R \simeq 0$
is reasonable in this regime, although its use is not strictly necessary
in developing the adiabatic limit \cite{c26}. However, the
extra algebra involved when $R \neq 0$ obscures rather than clarifies, so we will set
$R = 0$ for pedagogical reasons. Note also that $R \simeq 0$ 
is a valid approximation
for $T < 1$ MeV because weak collision effects are then unimportant ($D \simeq 0$ also
holds in this case).

With $R \simeq 0$, the QKEs simplify to
\begin{equation}
\frac{\partial}{\partial t}
\left( \begin{array}{c}
P_x \\ P_y \\ P_z
\end{array} \right) \simeq
\left( \begin{array}{ccc}
-D\ & -\lambda\ & 0 \\ \lambda\ & -D\ & -\beta \\ 0\ & \beta\ & 0
\end{array} \right)
\left( \begin{array}{c} P_x \\ P_y \\ P_z \end{array} \right) \,,
\label{R=0qkes}
\end{equation}
or, in a self-evident matrix notation,
\begin{equation}
\frac{\partial{\bf P}}{\partial t} \simeq {\cal K} {\bf P}\,.
\label{matrixqkes}
\end{equation}
We have dropped the explicit $(y,T)$ dependence for notational simplicity.

To solve equation (\ref{matrixqkes}), we first introduce the {\it
instantaneous diagonal basis} through
\begin{equation}
{\bf Q} = {\cal U} {\bf P}\,,
\end{equation}
where ${\cal U}$ is a time-dependent matrix that diagonalizes ${\cal K}$,
\begin{equation}
\label{kd}
{\cal K}_d \equiv {\rm diag}(k_1, k_2, k_3) = {\cal U} {\cal K} {\cal
U}^{-1}\,,
\end{equation}
with $k_{1,2,3}$ being eigenvalues.
In the instantaneous diagonal basis, equation (\ref{matrixqkes}) becomes
\begin{equation}
\frac{\partial {\bf Q}}{\partial t} \simeq {\cal K}_d {\bf Q} - 
{\cal U} \frac{\partial
{\cal U}}{\partial t}^{-1} {\bf Q}\,.
\label{udu}
\end{equation}
The {\it adiabatic limit} of equation (\ref{matrixqkes}) is defined by the
approximation
\begin{equation}
{\cal U} \frac{\partial {\cal U}}{\partial t}^{-1} \simeq 0 \,.
\end{equation}
Equation (\ref{udu}) is then formally solved to yield
\begin{equation}
{\cal P}(t) = {\cal U}^{-1}(t) e^{\int_0^t {\cal K}_d(t') dt'}
{\cal U}(0) {\cal P}(0) \,.
\label{adsoln}
\end{equation}
The time dependences of the various quantities have been reinstated for
obvious reasons.

Under most circumstances, for instance $|\lambda| \gg D$ and $|\beta| \gg
D$ individually, the eigenvalue spectrum consists of a complex conjugate
pair \cite{deviantearly,c25},
\begin{equation}
k_{1,2} \equiv - d \pm i \omega \,,
\end{equation}
together with a negative (real) eigenvalue,
\begin{equation}
k_3 = - \frac{\beta^2 D}{d^2 + \omega^2}\,.
\end{equation}
The easily derived relations,
\begin{eqnarray}
d & = & D + \frac{k_3}{2},\\
\omega^2 & = & \lambda^2 + \beta^2 + k_3 D + \frac{3}{4} k_3^2\,,
\end{eqnarray}
should also be noted. It is useful to call $d$ the 
{\it oscillation-affected decoherence function} and $\omega$ the 
{\it matter- and collision-affected oscillation frequency}. The very important
third eigenvalue $k_3$ is the relaxation rate for the $\nu_{\alpha}
\leftrightarrow \nu_s$ process (see below). For many
applications $|k_3|$ is small, so that $d \simeq D$ and $\omega \simeq 
\sqrt{\lambda^2 + \beta^2}$. In order to use equation (\ref{adsoln}), the 
diagonalization matrices must also be computed. The explicit expressions
can be found in Refs.\ \cite{deviantearly,c25}.

The adiabatic limit solution supplies a useful picture for the
dynamics (in its domain of applicability \cite{deviantearly} of
course). (Some of the following insights were obtained earlier using a
related approach \cite{c12,c17}.)  Consider first the case of zero
decoherence, $D = 0$. This situation obtains after neutrino
decoupling. The eigenvalues $k_i$ are then \cite{deviantearly,c25}
\begin{equation}
k_1 = k_2^* = i \sqrt{\beta^2 + \lambda^2}\,,\quad k_3 = 0\,,
\end{equation}
where $\omega = \sqrt{\beta^2 + \lambda^2}$ is in this case exactly
equal to the usual matter-affected oscillation frequency. The
adiabatic solution of equation (\ref{adsoln}) is then just a
restatement of the usual adiabatic matter-affected evolution one
obtains by solving the Schr\"{o}dinger Equation. For instance, one can
show that equation (\ref{adsoln}) becomes
\begin{equation}
P_z(t) = \cos 2\theta_m(t) \cos 2\theta_m(0)\,,
\end{equation}
with the initial conditions $P_{x,y}(0) = 0$ and $P_{z,0}(0) = 1$
(i.e.\ no sterile neutrinos initially). The function $\cos 2\theta_m
\equiv \lambda/\sqrt{\lambda^2 + \beta^2}$ is the cosine of twice the
standard matter-affected mixing angle. The oscillatory terms arising
from $k_{1,2}$ have been taken to average to zero. This is the
standard adiabatic MSW result. For instance if the effective matter
potential (and hence $\lambda$) varies from $+\infty$ at $t = 0$ to
$-\infty$ at $t = \infty$ through the resonance $\lambda = 0$, then
$P_z$ evolves from $1$ to $-1$, signalling complete $\nu_{\alpha} \to
\nu_s$ conversion.  A simple physical picture then results: as
neutrinos with momentum $y$ cross a narrow MSW resonance centered at
$y = y_{res}$, adiabaticity guarantees full $\nu_{\alpha}
\leftrightarrow \nu_s$ conversion. (Similarly antineutrinos, but note
that $\overline{y}_{res} \neq y_{res}$ when $L^{(\alpha)} \neq 0$.)
For a narrow resonance, the QKEs can be approximated by equations
which relate the rates of change of the distribution functions to the
speeds at which the resonance momenta move through the distributions
\cite{c13,c27}.  In practice, such an approximation can often also be
validly used for temperatures slightly higher than $1$ MeV.

Now consider the opposite case, where collisional effects
dominate. The real parts of $k_{1,2}$ then completely damp the
oscillatory behaviour driven by the imaginary parts, and a nonzero
$k_3$ enters the game. Equation (\ref{adsoln}) then allows one to
solve for the coherence function $P_{y}$ in terms of $P_z$, yielding
\begin{equation}
P_y(t) \simeq \frac{k_3(t)}{\beta(t)} P_z(t)\,.
\label{PyPz}
\end{equation}
(One can also solve for $P_x$.) Substituting for $P_y$ in equation
(\ref{R=0qkes}), one obtains
\begin{equation}
\frac{\partial P_z}{\partial t} \simeq k_3 P_z
\label{Pzeqn}
\end{equation}
as a self-contained differential equation for $P_z$ and hence for the
distribution functions. (A similar equation follows for
antineutrinos.)  Because the coherences $P_{x,y}$ have been
eliminated, one can speak of the {\it classical Boltzmann limit} of
the QKEs. Because $k_3$ is negative, and under most interesting
circumstances much smaller in magnitude than the decoherence rate $D$,
equation (\ref{Pzeqn}) describes a relatively slow collision-dominated
evolution of $P_z$ towards zero, that is, towards $N_{\nu_{\alpha}} =
N_{\nu_s}$. It can be viewed as a relaxation effect.

\subsubsection{Neutrino Asymmetry Evolution}

Neutrino asymmetries and chemical potentials are important quantities.
As previously explained, active-active oscillations are expected to
have strong effects only if $e$, $\mu$ and $\tau$ asymmetries are
unequal, and an $e$-like asymmetry directly affects BBN through the
proton-neutron interconversion reactions.  Furthermore, the effective
matter potential of equation (\ref{Valpha}) has a term proportional to
a linear combination of asymmetries. Large asymmetries therefore imply
small matter-affected mixing angles $\theta_m$, and hence suppressed
oscillations. This is of particular concern for sterile neutrino
models: sufficiently large asymmetries will suppress active to sterile
oscillations with important implications for the consistency of these
models with BBN.

Using the QKEs together with $\alpha + s$ lepton number conservation,
one can easily show that
\begin{equation}
\frac{d L_{\nu_{\alpha}}}{dt} =
\frac{1}{2 n_{\gamma}} \int \beta (P_y - \overline{P}_y) N^{eq}(p,0,T) dp 
\label{dLdt}
\end{equation}
is the evolution equation for $L_{\nu_{\alpha}}$ under the action of
$\nu_{\alpha} \leftrightarrow \nu_s$ and $\overline{\nu}_{\alpha}
\leftrightarrow \overline{\nu}_s$ oscillations. We will study this
simple two-flavor toy model before considering more realistic
scenarios.

\begin{figure}
\begin{center}
\includegraphics[angle=270,scale=0.5,viewport=0 0 800 100]{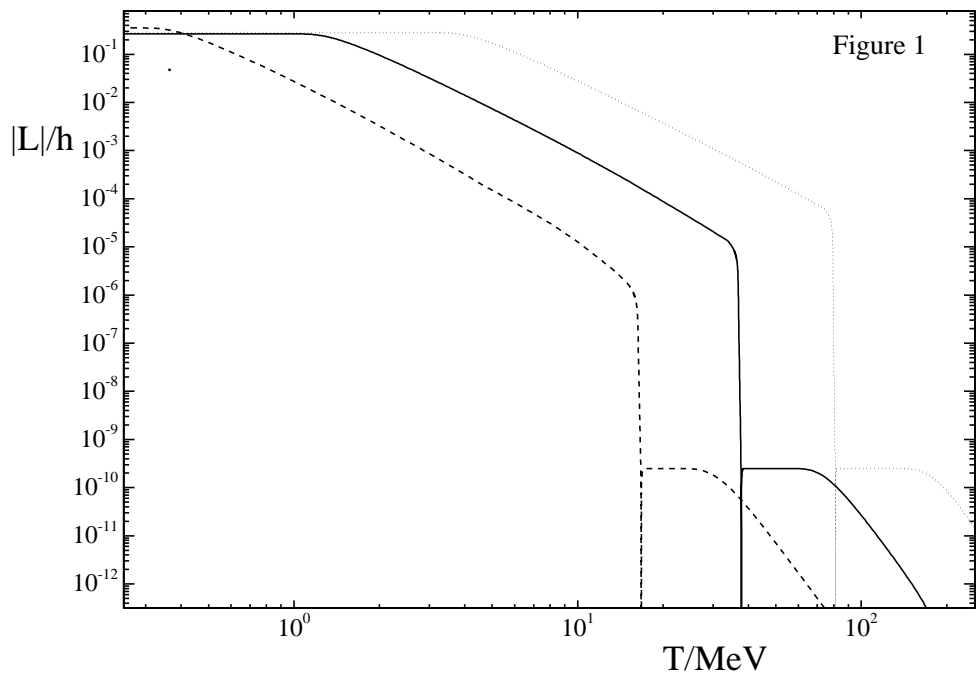}
\end{center}
\caption{Neutrino asymmetry growth curves driven by $\nu_\tau
\leftrightarrow \nu_s$ and $\overline{\nu}_\tau \leftrightarrow 
\overline{\nu}_s$ oscillations. The mixing angle is selected to be 
$\sin^2 2\theta_0 = 10^{-8}$. The three curves correspond to
$\Delta m^2 = -0.5$, $-50$ and $-5000$ eV$^2$, reading from left to
right. This figure is taken from Ref. \cite{c18}.}
\label{cosmofig1}
\end{figure}

The QKEs can be numerically integrated to yield asymmetry evolution
curves.  (For numerical reasons, it is useful to include equation
(\ref{dLdt}) in the coupled system, even though it is redundant).  Let
us suppose that all neutrino asymmetries start out being small,
perhaps of the order of the observed baryon asymmetry $\sim
10^{-10}$. There is no guarantee that this is realistic, but up to now
most of the interesting work in the literature has focussed on this
case.  When $\Delta m^2 < 0$ and the vacuum mixing angle is small,
there is a large range of parameters for which explosive asymmetry
growth begins at a certain critical temperature $T_c$, as displayed in
Figure\ \ref{cosmofig1}. This is a runaway positive feedback phenomenon
that occurs when an $L^{(\alpha)} = 0$ approximate fixed point changes
from being stable to unstable at $T = T_c$ \cite{c12,c17}.  After a
short spurt of quasi-exponential growth, the evolution settles into a
less dramatic but still significant approximate power law phase,
before reaching a steady state plateau at a value in the range $0.2 -
0.35$ (depending on the oscillation parameter choice)
\cite{c13}.\footnote{Reference \cite{dhps} found much smaller values
for the final asymmetries. A critique of this work can be found in
Ref.\ \cite{c28}. See also Refs.\ \cite{sorri,burassemikoz}.}  The
baryon asymmetry in the plasma, which couples in through the
$\eta$-term in $L^{(\alpha)}$, seeds the neutrino asymmetry
growth. The existence of the asymmetry growth effect must be carefully
taken into account when calculating sterile neutrino production for
BBN purposes.

The main features of the asymmetry growth curves have been understood,
and are discussed in some detail in Refs.\ \cite{c12,c13,c17,c28}.  In
a nutshell, the explosive growth phase is governed by the collision
dominated dynamics leading to equations (\ref{PyPz}) and
(\ref{Pzeqn}), while the approximate power law regime has adiabatic
MSW transitions as the driving force.  It has been shown that
asymmetry growth resembling that shown in Figure\ \ref{cosmofig1} will
occur for the parameter space region
\begin{eqnarray}
&\Delta m^2 < 0\quad {\rm with}\quad |\Delta m^2| 
\stackrel{>}{\sim} 10^{-4}\ {\rm eV}^2 \,, &
\nonumber\\
& 10^{-10} \stackrel{<}{\sim} \sin^2 2\theta_0 \stackrel{<}{\sim} 
{\rm few} \times 10^{-5} 
\left( \frac{ {\rm eV}^2 }{ |\Delta m^2| } \right)^{1/2}\,.&
\label{parspace}
\end{eqnarray}
For $|\Delta m^2| < 10^{-4}$ eV$^2$, asymmetry can be generated, but
it tends to be oscillatory \cite{c7,c20,c23prime,c23}.  Values of
$\sin^2 2 \theta_0$ smaller than $10^{-10}$ are too small to generate
an asymmetry, while values exceeding the upper bound above lead to
copious sterile neutrino production which is also detrimental to
asymmetry growth. One can also show that the critical temperature
$T_c$ is roughly given by
\begin{equation}
T_c \sim (16 \to 20\ {\rm MeV}) \left( 
\frac{|\Delta m^2| \cos 2\theta_0}{{\rm eV}^2} \right)^{1/6}\,.
\end{equation}

\subsection{Active-Sterile Oscillations: Applications}

\subsubsection{Sterile Neutrino Production}

For the moment, let us continue to study the toy universe where only
$\nu_{\alpha} \leftrightarrow \nu_s$ and $\overline{\nu}_{\alpha}
\leftrightarrow \overline{\nu}_s$ oscillations occur (for a particular
$\alpha$). An interesting question is: for a given choice of $\Delta
m^2$ and $\theta_0$, how does sterile neutrino production affect BBN?
For $\alpha = \mu$ or $\tau$ the effect is restricted to potentially
increasing the energy density in relativistic species and hence the
expansion rate of the universe.  (Note that for $\Delta m^2
\stackrel{<}{\sim} 10^{-7}$ eV$^2$ significant $\nu_{\mu,\tau}
\leftrightarrow \nu_s$ oscillations can occur only after neutrino
decoupling, so the overall energy density is unaffected.) For $\alpha
= e$, the possible effects are more complicated.  Because of the
reactions in equation (\ref{npreactions}), the possible existence of
an $e$-like asymmetry must be carefully considered, and one has to
take into account the effect of distortion and depletion of the
$\nu_e$ spectrum below the decoupling temperature
\cite{c7,c23prime,c23}.

The detailed results will clearly depend on whether or not the plasma
has large neutrino asymmetries. As we have just seen, an $\alpha$-like
asymmetry will be generated during the $T \sim 10$'s of MeV epoch by
oscillations if the parameters obey equation (\ref{parspace}). Other
mechanisms, such as Affleck-Dine, can create large asymmetries at much
higher temperatures.  Unfortunately, prospects for directly detecting
the relic neutrino sea remain remote, so there is much room for
theoretical speculation.

The important pioneering works on sterile neutrino production were
performed before the asymmetry generation effect was understood
\cite{c29,c30,c31,c32,c33,c34}.  By neglecting the baryon asymmetry
and setting all neutrino asymmetries to zero, one obtains a simplified
system driven by identical neutrino and antineutrino oscillation
dynamics which are decoupled from each other. For $\Delta m^2 > 0$,
this simplification is consistent. However, for $\Delta m^2 < 0$, the
$L^{(\alpha)} = 0$ approximate fixed point becomes unstable and the
zero asymmetry constraint is not justified.

A useful summary of the pioneering results is contained in Figures\ 2
and 4 from Ref.\ \cite{c32}.  These plots show contours in the $\Delta
m^2 - \sin^2 2\theta_0$ plane corresponding to different values of
$g_{max}$, a parameter which quantifies sterile neutrino contribution
to the expansion rate, equation (\ref{H}), through an effective
increase in the number of fermionic degrees of freedom contributing to
$g_*$. (We will call this parameter $\Delta N_{eff}$ from now on.  Note
that for $\alpha = e$ its meaning is more subtle \cite{c32}.) The
precise constraint one should use is at this stage unclear, because of
uncertainties in the primordial light element abundance observations.
Nevertheless, these plots provide interesting information for neutrino
model builders. For instance, a $\nu_{\mu} \to \nu_s$ solution to the
atmospheric neutrino problem with $\Delta m^2 \sim 10^{-2.5}$ eV$^2$
and $\sin^2 2\theta_0 \sim 1$ would appear to go hand-in-hand with a
fully thermalized $\nu_s$ prior to BBN. Such a situation would be
cosmologically disfavored because of concomitant $^4$He
overproduction.

As already noted, for $\Delta m^2 < 0$ one has to take into account
the asymmetry generation that occurs simultaneously with sterile
neutrino production. The qualitative effect is clear: because
asymmetries suppress oscillations, the rigorous $\Delta m^2 < 0$
bounds should be weaker than those displayed in Figure\ 4 of Ref.\
\cite{c32}, for example.  However, detailed calculations
\cite{c17,c35} reveal that the asymmetry amplification process does
not modify the ``traditional'' bounds by very much, except when the
BBN constraint on $\Delta N_{eff}$ is taken to be rather stringent,
say $\Delta N_{eff} < 0.1$. The reason for this is that sterile
neutrino production tends to delay the onset of asymmetry growth (see
Figure\ 12 of Ref.\ \cite{c35} for a clear illustration).

All of the above related to the artificial two-flavor
$\nu_{\alpha}/\nu_s$ case. But the most dramatic consequence of
asymmetry creation occurs in more realistic multiflavor
situations. This is simply because an asymmetry created by one
active-sterile oscillation mode can suppress a {\it different}
active-sterile mode \cite{c12,c17,c18,c35}.  The most studied case was
motivated by the atmospheric neutrino problem. As noted above, the
$\nu_{\mu} \to \nu_s$ solution to the atmospheric neutrino problem
appears at first sight to be cosmologically disfavored. However, tiny
mixing between the $\nu_s$ and a more massive $\nu_{\tau}$ can
completely change the dynamics, because the nonzero $L_{\nu_{\tau}}$
created by the small-angle, negative $\Delta m^2$ $\nu_{\tau} \to
\nu_s$ mode can suppress the $\nu_{\mu} \to \nu_s$ oscillations that
would be very strong otherwise. Explicit calculations show that if the
mass eigenvalues associated with the $\nu_{\mu}/\nu_s$ subsystem are
much less than an eV, then a $\nu_\tau$ mass at the few eV level or
higher is required \cite{c12,c17,c18,c35}.  (Though somewhat large
values were obtained in Ref.\ \cite{c36}, the later independent
results of Ref.\ \cite{c35} agree with the original results of Refs.\
\cite{c12,c17}.)

\subsubsection{$e$-Like Asymmetry and $^4$He Abundance.}

Suppose a small-angle active-sterile mode with a relatively large but
negative $\Delta m^2$ creates a large asymmetry of a certain flavor
fairly early. Later on, it is certainly possible for other oscillation
modes to reprocess the flavor of the asymmetry. For instance,
small-angle $\nu_{\tau} \to \nu_s$ oscillations might create a large
$L_{\nu_{\tau}} \sim 0.3$ which gets reprocessed into a reasonably
large (say $\sim 0.01$) $L_{\nu_e}$ by $\nu_{\tau} \leftrightarrow
\nu_e$ oscillations. Recall that the usually innocuous active-active
modes can become important after flavor asymmetries get created.

Several scenarios of this type have been investigated in the
literature \cite{c11,c13,c27,c37}.  One has to study the peculiarities
of each neutrino model on a case-by-case basis. The more flavors one
has, the more complicated the analysis becomes. The most ambitious
attempt to date relates to the mirror matter model with three active
and three mirror neutrino flavors \cite{c11}.

By way of example, Ref.\ \cite{c13} considered a model with the three
active neutrinos and one light sterile flavor with the mass hierarchy
$m_{\nu_\tau} \gg m_{\nu_\mu}, m_{\nu_e}, m_{\nu_s}$.  A $\tau$-flavor
asymmetry is first created by the $\nu_\tau \leftrightarrow \nu_s$
mode, and subsequently reprocessed into $e$- and $\mu$-like
asymmetries by $\nu_\tau \leftrightarrow \nu_e$ and $\nu_\tau
\leftrightarrow \nu_\mu$ oscillations. The $e$-like asymmetry is of
most interest because of its role in the neutron-proton
interconversion reactions. Final values for $L_{\nu_e}$ around $0.02
(T_{\nu_e}/T_{\gamma})^3$ for $\Delta M^2 > 10$ eV$^2$ were found,
where $\Delta M^2$ is the squared-mass splitting between $\nu_\tau$
and the lighter flavors.  The effect of such an asymmetry on $^4$He
abundance is roughly equivalent to {\it reducing} the expansion rate
by, effectively, half of a neutrino: $\Delta N_{eff} \sim -0.5$.  Note
that the sign of the asymmetry is crucial here: positive $L_{\nu_e}$'s
reduce the $^4$He yield (equivalent to reducing the expansion rate),
while negative values increase the yield (equivalent to increasing the
expansion rate).  The sign of $L_{\nu_e}$ is controlled by the sign of
the $L_{\nu_\tau}$ originally created, which in turn is controlled by
the unknown initial (high $T$) values of the asymmetries. It turns out
that the results quoted above assume initial conditions that lead to a
positive $L_{\nu_{\tau}}$.

Clearly, the relevance or otherwise of these studies for understanding
nature hinges on the as yet unclear existential status of light
sterile neutrinos. Future terrestrial experiments such as SNO,
MiniBOONE and the long baseline projects will help to clarify the
situation. Better information about the mass and mixing angle spectrum
for neutrinos will be needed, as well as better information about
primordial light element abundances. Precision cosmic microwave
background measurements also have a role to play by helping to pin
down the baryon density (a crucial input into BBN calculations), as
well as the expansion rate of the universe at the time of photon
decoupling \cite{lange01,balbi00,di2001}.

\subsection{Remarks About Active-Active Oscillations}

Collision-affected active-active oscillations have received very
little attention because the relevant Quantum Kinetic Equations are
rather complicated, as can be seen by comparing equation (\ref{4})
with equation (\ref{5}). The studies that have been done typically
focussed on collision-unaffected adiabatic transitions, for which a
simpler treatment is possible. The works alluded to in the previous
subsubsection in fact used such an approach.  The collision-affected
case remains largely virgin territory: the required QKEs have been
written down \cite{MandT,raff2}, but remain unsolved, except for a
certain interesting special case \cite{bvs,bvs2}.  The special
situation is that of propagation through a ``flavor-blind''
medium. Examples include an active-active oscillating system
traversing a dense nucleonic medium, and a $\nu_{\mu}/\nu_{\tau}$
system propagating through an electron-positron plasma. In both cases,
the microscopic collisions do not distinguish between the flavors of
the oscillating system. While the practical relevance of these systems
is not entirely clear, a very interesting phenomenon has been
discovered: synchronization of modes in the rapid collision limit. In
vacuum, and in a refractive medium, neutrino oscillation lengths
generally depend on energy. It has been found that rapid flavor-blind
collisions force all modes to have the same oscillation length, which
has been analytically proven to be a kind of thermal average of the
oscillation lengths the neutrinos would have in the absence of the
collisions.  This leads at a sharpening of MSW transitions, and it
reflects an interesting persistence-of-quantal-coherence effect which
may find application outside of the neutrino domain \cite{bvs}.

\section{NEUTRINOS IN SUPERNOVAE AND PROTO-NEUTRON STARS}

The current gravitational core collapse supernova paradigm is based
on the original suggestion by Colgate and White~\cite{CW66} that the
explosion derives from neutrino energy deposition.  However, since
that time, much has been learned about neutrino--matter interactions
that have modified the original model.  It has also been realized that
the implementation of accurate neutrino transport is critical, since
the deposited energy is a small fraction ($\le1\%$) of the available
gravitational binding energy of the remnant
($\sim3GM^2/5R\approx3\cdot10^{53}(M/1.4{\rm~M}_\odot)^2$ ergs).

The initial phases of the supernova begin with the destabilization and
collapse of the core of a massive star (see Ref.~\cite{BBAL} for 
the important physics which occurs during this period).  The
collapsing core, which is initially composed of iron peak elements
with a net electron content $Y_e=n_e/n_B\simeq0.41-0.43$, divides into
two portions: an inner, homologous (with infall velocity roughly
proportional to the radius) core and an outer region that collapses
supersonically.  The infalling matter maintains a nearly constant
entropy per baryon $s\approx1$; the entropy gain from
out-of-equilibrium weak interactions is balanced by the energy loss
from escaping neutrinos.  During the collapse, electrons and protons
are converted into neutrons and neutrinos as the matter attempts to
maintain beta equilibrium.  When the central density reaches about
$10^{12}$ g cm$^{-3}$, neutrinos are unable to escape on dynamical
timescales and are essentially  frozen, or `trapped', in the matter.
The lepton number thereafter remains fixed, at a value of 
$Y_L=Y_e+Y_{\nu_e}\simeq0.4$.

The collapse continues until the central density exceeds $n_0$ when
the increased pressure from strong interactions reverses it.  A shock
is formed at the outer edge of the inner core (where sound waves from
the center accumulate) and begins to propagate through the infalling
outer region of the core.  But this shock, called the ``bounce''
shock, stalls at a distance from the center of about 150 km, due to
the energy expended in nuclear dissociation and neutrino losses.  The
stalled shock becomes an accretion shock which separates
supersonically infalling matter from hot matter slowly settling onto
the inner core, a PNS.  After 10--20 ms, the overall
structure evolves quasi-hydrostatically.  The nucleons produced by the
dissociation of heavy nuclei are heated by neutrino absorption from
the hot, newly formed PNS.  Because neutrino emission
varies as $T^6$, there is a point, known as the ``gain
radius'', at which the heating exceeds the cooling.

Current calculations differ to some extent as to the outcome of this
scenario, due to varying input physics and level of approximation to
the neutrino transport problem.  In addition, the negative entropy
gradient that naturally exists is unstable against convection.  This
convection, which is neutrino-driven, seems to eventually assist in
reviving the shock in some models.  However, convection can only be
realistically modeled in three dimensions, a task that is only
beginning to be addressed.  Additional fluid instabilities may also
arise near the neutrinosphere, at which the neutrino optical depth to
infinity is of order unity and where the neutrinos can begin to freely
escape the star.  The shock position during this hydrostatic epoch is
determined by a delicate balance between thermal pressure caused by
neutrino heating and the ram pressure of infalling matter~\cite{BG93}.
A successful supernova results if this balance becomes unstable, which
could occur if the accretion shock can be maintained at a sufficient
distance for a long enough time.  The ram pressure decreases as
material from less dense regions of the outer core is encountered, and
could be eventually overcome by the more steady neutrino radiation
from the core.  Nevertheless, explosions in model simulations, even
when they occur, appear to be marginal at best.  Aside from
uncertainties stemming from simulating 3-D, general relativisitic,
neutrino transport, the initial structure of the pre-collapse core, the
equation of state, and neutrino opacities and emissivities, all play
roles in the outcome \cite{bnat}.

\subsection{Role of Neutrinos in Gravitational Collapse Supernovae}

Neutrino transport in the supernova environment is described by a
Boltzmann tranport equation, derivable from the kinetic equation
(\ref{4}) by retaining only the diagonal elements in the density matrix 
$\rho({\bf r},{\bf r}^\prime)$.   Even with this simplification, 
it is a nonlinear integro-partial differential equation
that describes the time rate of change of the neutrino distribution
function $f$.  Advances made to date in the numerical solution of 
this equation in the supernova context may be found 
in Refs.~\cite{Trans}.  Historically, multigroup methods (in
which the equation is discretized in energy groups) have involved
moment expansions.  When the temporal derivative of the first moment
of $f$ is set to zero, a diffusion equation is obtained, but this
cannot adequately handle the free-streaming regime at low densities.
Flux limiting schemes have been used to bridge the diffusive and
free-streaming regimes, but these are somewhat arbitrary and accurate
calibration depends upon neutrino opacities and dynamics.  In
addition, there is a problem with the coupling of different
neutrino-energy groups, especially because of $\nu-e$ scatterings,
which involve large energy transfers.  An additional complication in
supernovae is that the approach to thermal and chemical
equilibrium, and the conversion of diffusive flow to free streaming,
occur simultaneously in space and time.  Even with modern parallel
supercomputers, it is necessary to integrate the Boltzmann equation
over solid angles to reduce the dimensionality of the problem, with a
corresponding loss of information about the neutrino angular
distribution function.  This could be important in regimes in which
neutrino-driven convection, a 3-D phenomenon, is occuring.

Crucial weak interaction processes in the supernova environment include
\begin{equation}
p + e^- \rightarrow n + \nu_e,\quad (A,Z) + e^- \rightarrow (A,Z-1) + 
\nu_e\,,  
\label{neu}
\end{equation}
\begin{equation}
\nu + (A,Z) \rightarrow \nu + (A,Z) \,,  
\label{coh}
\end{equation}
\begin{equation}
\nu + e^- \rightarrow \nu + e^-, \qquad
\nu + (A,Z) \rightarrow \nu + (A,Z)^*\,, 
\label{scat}
\end{equation}
\begin{eqnarray}
e^+ + e^- \rightarrow  \nu + \bar{\nu}, \quad
(A,Z)^* \rightarrow  (A,Z) + \nu + \bar{\nu}, \quad
(n,p)\rightarrow(n,p)+\nu+\bar\nu\,. 
\label{pair}
\end{eqnarray}
Reactions (\ref{neu}) begin the process of neutronization and decrease of
$Y_L$, whose value after trapping determines the masses of the
homologous core and initial PNS, and thus the available
energy for the shock and subsequent neutrino emissions.  The equation
of state also influences these quantities, most importantly through
the nuclear symmetry energy.

In the subnuclear density regime, the coherent scattering reaction
(\ref{coh}) from nuclei in a lattice is the most important opacity
source (see Section 3.2).
The reactions (\ref{scat}) are important in changing the neutrino
energy, and in achieving thermodynamic equilibrium.  As referred to in
Section 5.1, the large energy transfers of these processes is a hurdle
for numerical calculations.  The reactions (\ref{pair}) are also
important in achieving thermodynamic equilibrium.  The bremsstrahlung
($n+n\rightarrow n+n+\nu+\bar\nu$) and modified Urca ($n+p\rightarrow
n+n+e^++\nu+\bar\nu$) processes involving nucleons dominate in many
circumstances.  For example, the production and thermalization of
$\mu$ and $\tau$ neutrinos, which receives contributions from all the
reactions (\ref{pair}), is dominated by nucleon bremsstrahlung for
$n>0.005$ and $T<15$ MeV \cite{TBH}.  The modified Urca process
dominates the cooling of PNSs if direct Urca processes
involving nucleons, hyperons or other strange particles do not occur.

\subsection{Neutrinos From Proto-Neutron Stars}

A PNS is born in the aftermath of the gravitational
collapse of the core of a massive star accompanying a
successful supernova explosion.  During the first tens of seconds of
evolution, nearly all ($\sim$ 99\%) of the remnant's binding energy is
radiated away in neutrinos of all flavors
\cite{burr86,KJ95,burr99a,pons99,Pon00a,Pon01b}.  The neutrino luminosities
and the emission timescale are controlled by several factors, such
as the total mass of the PNS and the opacity at supranuclear
density, which depends on the composition and EOS.
One of the chief objectives in modeling PNSs is to infer their
internal compositions from  neutrino signals detected from future
supernovae like SuperK, SNO and others under consideration, including
UNO \cite{AIP}.

\subsubsection{General Description of the Birth of Proto-Neutron Stars}

\begin{figure}
\begin{center}
\includegraphics[angle=90,scale=0.6]{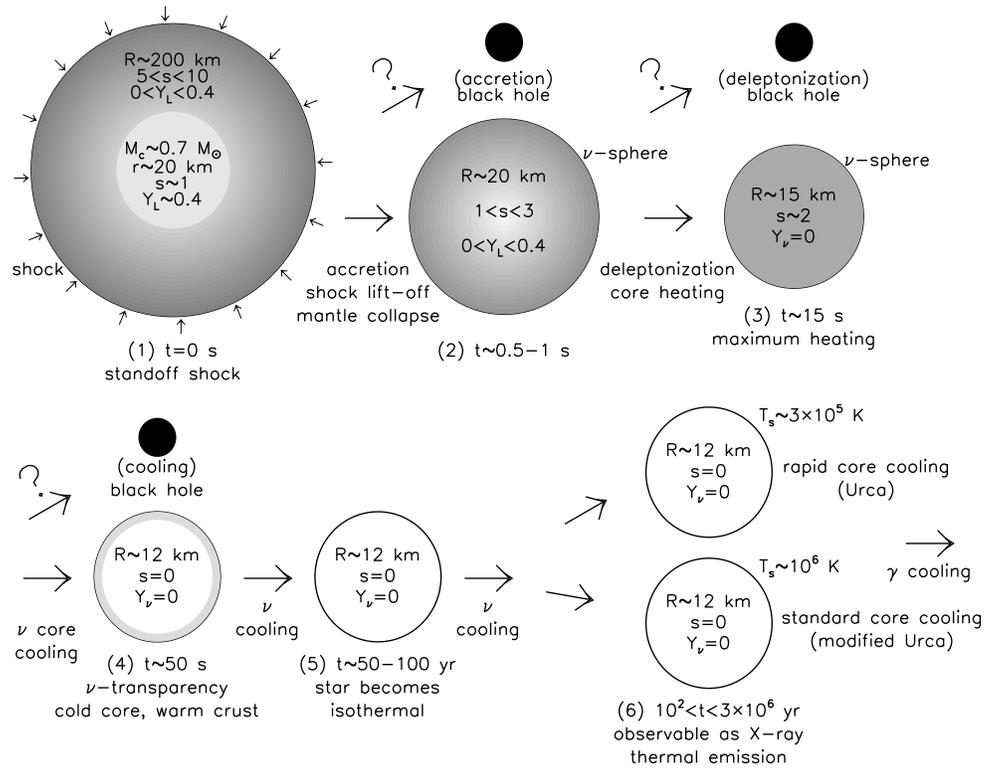}
\end{center}
\caption{The main stages of evolution of a neutron star.  Shading
indicates, approximately, relative temperatures.  This figure is from
Ref. \cite{prak97a}.}
\label{pict1}
\end{figure}

The evolution of a PNS proceeds through several distinct stages
\cite{burr86,supernova} and with various outcomes~\cite{prak97a}, as
shown schematically in Figure~\ref{pict1}. Immediately following core
bounce and the passage of a shock through the outer PNS's mantle, the
star contains an unshocked, low entropy core of mass $\simeq0.7$
M$_\odot$ in which neutrinos are trapped (stage 1 in the figure). The
core is surrounded by a low density, high entropy ($5<s<10$) mantle
that is both accreting matter from the outer iron core falling through
the shock and also rapidly losing energy due to electron captures and
thermal neutrino emission. The mantle extends up to the shock, which
is temporarily stalled about 200 km from the center prior to an
eventual explosion.

After a few seconds (stage 2), accretion becomes less important if the
supernova is successful and the shock has ejected the stellar envelope.
Extensive neutrino losses and deleptonization will have led to a loss
of lepton pressure and the collapse of the mantle.  If enough
accretion has occurred, however, the star's mass could increase beyond the
maximum mass capable of being supported by the hot, lepton-rich
matter.  If this occurs, the remnant collapses to form a black hole
and its neutrino emission is believed to quickly cease \cite{burr88}.

Neutrino diffusion deleptonizes the core on time scales of 10--15 s
(stage 3).  Diffusion time scales are proportional to
$R^2(c\lambda_\nu)^{-1}$, where $R$ is the star's radius and
$\lambda_\nu$ is the effective neutrino mean free path.  This generic
relation illustrates how both the EOS and the composition influence
evolutionary time scales.  The diffusion of high-energy (200--300 MeV)
$\nu$s from the core to the surface where they escape as low-energy
(10--20 MeV) $\nu$s generates heat (a process akin to joule
heating). The core's entropy approximately doubles, producing
temperatures in the range of 30--60 MeV during this time, even as
neutrinos continue to be prodiguously emitted from the star's effective
surface, or $\nu-$sphere.

Strange matter, in the form of hyperons, a Bose condensate, or quark
matter, suppressed when neutrinos are trapped, could appear
at the end of the deleptonization.  Its appearance would
lead to a decrease in the maximum mass that matter is capable of
supporting, implying metastability of the neutron star and another
chance for black hole formation~\cite{prak97a}.  This would occur
if the PNS's mass, which must be less than the maximum mass of hot,
lepton-rich matter (or else a black hole would already have formed),
is greater than the maximum mass of hot, lepton-poor matter.  However,
if strangeness does not appear, the maximum mass instead increases
during deleptonization and the appearance of a black hole would be
unlikely unless accretion in this stage remains significant.

The PNS is now lepton-poor, but it is still hot.  While the star has
zero net neutrino number, thermally produced neutrino pairs of all
flavors dominate the emission.  The average neutrino energy slowly
decreases, and the neutrino mean free path increases.  After
approximately 50 seconds (stage 4), $\lambda\simeq R$, and the star
finally becomes transparent to neutrinos.  Since the threshold density
for the appearance of strange matter decreases with decreasing
temperature, a delayed collapse to a black hole is still possible
during this epoch.

Following the onset of neutrino transparency, the core continues to
cool by neutrino emission, but the star's crust remains warm and cools
less quickly. The crust is an insulating blanket which prevents the
star from coming to complete thermal equilibrium and keeps the surface
relatively warm ($T\approx3\times10^6$ K) for up to 100 years (stage
5).  The temperature of the surface after the interior of the
star becomes isothermal (stage 6) is determined by the rate of
neutrino emission in the star's core and the composition of the surface.

\subsubsection{The Proto-Neutron Star Evolution Equations}

The equations that govern the transport of energy and lepton number in
a PNS are obtained from the Boltzmann equation for
massless particles \cite{burr86,lind66,thor81,pons99}.  We will focus
on the non-magnetic, spherically symmetric situation, and note that fluid
velocities are small enough so that hydrostatic equilibrium is nearly
fulfilled.  Under these conditions, the neutrino transport equations
in a stationary metric
\begin{eqnarray}
\label{metric}
ds^2=-e^{2\phi}dt^2+e^{2\Lambda}dr^2+r^2d\theta^2+r^2\sin^2\theta \,d\Phi^2\,
\end{eqnarray}
are:
\begin{eqnarray}
\label{number}
\frac{\partial ({N_{\nu}/n_B})}{\partial t} &+&
 {\frac{\partial (e^{\phi} 4 \pi r^2 F_{\nu})}{\partial a}}
= e^\phi \frac{S_N}{n_B} \\
\label{energy}
\frac{\partial ({J_{\nu}/n_B})}{\partial t} &+& P_{\nu} \frac{\partial
({1/n_B})}{\partial t} + e^{-\phi} {\frac{\partial (e^{2 \phi} 4 \pi
r^2 H_{\nu})}{\partial a}} = e^\phi \frac{S_E}{n_B} \,,
\end{eqnarray}
where $n_B$ is the baryon number density and $a$ is the enclosed baryon number
inside a sphere of radius $r$. The quantities 
$N_{\nu}$, $F_{\nu}$, and $S_N$ are the number density, number flux and number
source term, respectively, while $J_{\nu}$, $H_{\nu}$, $P_{\nu}$, and $S_E$ are
the neutrino energy density, energy flux, pressure, and the energy
source term, respectively.

In the absence of accretion,   the enclosed baryon number
$a$ is a convenient Lagrangian variable.  The equations to be solved
split naturally into a transport part, which has a strong time
dependence, and a structure part, in which evolution is much slower.
Explicitly, the structure equations are
\begin{eqnarray}
\label{a-struc}
{{\partial r}\over{\partial a}} = \frac{1}{4 \pi r^2 n_B e^{\Lambda}}
&,& \quad
\frac{\partial m}{\partial a} = \frac{\rho}{n_B e^{\Lambda}} \\
\label{b-struc}
\frac{\partial \phi}{\partial a} = \frac{e^{\Lambda}}{4\pi r^4 n_B}
{\left( m + 4\pi r^3 P \right)} &,& \quad
\frac{\partial P}{\partial a} = - (\rho + P)
\frac{e^{\Lambda}}{4\pi r^4 n_B}
{\left( m + 4\pi r^3 P \right)} \,.
\end{eqnarray}
The quantities $m$ (enclosed gravitational mass), $\rho$ 
(mass-energy density), and $P$ (pressure) include
contributions from the leptons.  To
obtain the equations employed in the transport, 
equation~(\ref{number}) may be combined with 
the corresponding equation for the electron
fraction
\begin{eqnarray}
\frac{\partial Y_e}{\partial t}=-e^\phi\frac{S_N}{n_B} \
\end{eqnarray} to obtain \begin{eqnarray}
\label{a-number}
\frac{\partial Y_L}{\partial t} +
e^{-\phi} {\frac{\partial (e^{\phi} 4 \pi r^2 F_{\nu})}{\partial a}}
= 0 \,.
\end{eqnarray}
Similarly, equation~(\ref{energy}) may be combined with the matter energy equation
\begin{eqnarray}
\frac {dU}{dt} + P \frac{d({1/n_B})}{dt} = - e^\phi \frac{S_E}{n_B} \,,
\end{eqnarray}
where $U$ is the specific internal energy.
The first law of thermodynamics yields
\begin{eqnarray}
e^\phi T\frac{\partial s}{\partial t} + e^\phi\mu_{\nu} \frac{\partial
Y_L}{\partial t} + e^{-\phi} {\frac{\partial e^{2 \phi} 4 \pi r^2
H_{\nu}}{\partial a}} = 0 \,.
\label{a-energy}
\end{eqnarray}

At high density and for $T\gg1$ MeV, the source
terms in the Boltzmann equation are sufficiently strong to ensure that
neutrinos are in thermal and chemical equilibrium with 
matter. Thus, the neutrino distribution function in these regions is
both nearly Fermi-Dirac and isotropic.  We can approximate the
distribution function as an expansion in terms of Legendre polynomials
to $O(\mu)$, which is known as the diffusion approximation.
Explicitly,
\begin{eqnarray}
f(\omega,\mu)= f_0(\omega) +  \mu f_1(\omega) \,, \quad
f_0 = [1+e^{\left(\frac{\omega-\mu_\nu}{kT}\right)}]^{-1} \,,
\end{eqnarray}
where $f_0$ is the Fermi--Dirac distribution function at equilibrium
($T=T_{mat}$, $\mu_{\nu}=\mu_{\nu}^{eq}$), with
$\omega$ and $\mu_\nu$ being the neutrino energy and chemical potential,
respectively.
In the diffusion approximation, $f_1(f_0)$ becomes \cite{pons99} 
\begin{eqnarray}
\label{f1}
f_1 = - D(\omega)
\left[ {e^{-\Lambda}} \frac{\partial f_0}{\partial r}
- {\omega} {e^{-\Lambda} \frac{\partial \phi}{\partial r}}
{\frac{\partial f_0}{\partial \omega}} \right] \,.
\end{eqnarray}
\noindent
The explicit form of the diffusion coefficient $D$ is given by
\begin{eqnarray}
D(\omega) = {\left( j+\frac{1}{\lambda_a}+\kappa^s_1 \right)}^{-1} \,.
\end{eqnarray}
The quantity $j=j_a+j_s$, where $j_a$ is the emissivity and 
$j_s$ is the scattering contribution to the source term. 
The absorptivity is denoted by 
$\lambda_a$ and $\kappa_1^s$ is the scattering contribution to the 
transport opacity.  
Substituting 
\begin{eqnarray}
\frac{\partial f_0}{\partial r} =
-\left( T \frac{\partial \eta_{\nu}}{\partial r} + \frac{\omega}{T}
\frac{\partial T}{\partial r}\right)
\frac{\partial f_0}{\partial \omega}~,
\end{eqnarray}
where $\eta=\mu_{\nu}/T $ is the neutrino degeneracy parameter, 
in equation~(\ref{f1}), yields
\begin{eqnarray}
f_1 = - D(\omega) e^{-\Lambda} \left[
T \frac{\partial \eta}{\partial r}
+ \frac{\omega}{T e^{\phi}} \frac{\partial (T e^{\phi})}{\partial r} \right]
\left(- \, \frac{\partial f_0}{\partial \omega} \right)\,.
\end{eqnarray}
Thus, the energy-integrated lepton and energy fluxes are
\begin{eqnarray}
F_{\nu}&=&- \, \frac{e^{-\Lambda} e^{-\phi}T^2}{6 \pi^2}
\left[ D_3 \frac{\partial (T e^{\phi})}{\partial r} +
(T e^{\phi}) D_2 \frac{\partial \eta}{\partial r}  \right] \nonumber \\
H_{\nu}&=&- \, \frac{e^{-\Lambda} e^{-\phi}T^3}{6 \pi^2}
\left[ D_4 \frac{\partial (T e^{\phi})}{\partial r} +
(T e^{\phi}) D_3 \frac{\partial \eta}{\partial r}  \right] \,.
\label{fluxes}
\end{eqnarray}
The coefficients $D_2$, $D_3$, and $D_4$ are defined by
\begin{eqnarray}
D_n = \int_0^\infty dx~x^n D(\omega)f_0(\omega)(1-f_0(\omega))~,
\label{d2d3}
\end{eqnarray} 
where $x=\omega/T$.  These diffusion coefficients 
depend only on the
microphysics of the neutrino-matter interactions. 
The fluxes appearing in the above equations are for one
particle species. To include all six neutrino types, we redefine the
diffusion coefficients in equation~(\ref{fluxes}):
\begin{eqnarray}
D_2=D_2^{\nu_e}+D_2^{\bar{\nu}_e}\,, \quad
D_3=D_3^{\nu_e}-D_3^{\bar{\nu}_e}\,, \quad
D_4=D_4^{\nu_e}+D_4^{\bar{\nu}_e}+4 D_4^{\nu_\mu}\,.
\end{eqnarray}

\subsubsection{Neutrino Luminosity from Proto-Neutron Stars}

A fair representation of the signal in a terrestrial detector can be
found from the time dependence of the total neutrino luminosity and
average neutrino energy together with an assumption of a Fermi-Dirac
spectrum with zero chemical potential.  The total neutrino luminosity
is globally the time rate of change of the star's gravitational mass, 
and due to energy conservation, is also
\begin{eqnarray}
L_\nu=e^{2 \phi} 4 \pi r^2 H_{\nu}\,
\end{eqnarray}
at the edge of the star.  However, since the spectrum is not precisely
Fermi-Dirac at the neutrinosphere, a diffusion scheme only
approximates the average energy.  The average energy can be
approximated as $<E_\nu>\approx3T_\nu$, where $T_\nu$ is a
mass-averaged temperature in the outermost zone.

\begin{figure}[hbt]
\includegraphics[scale=0.7]{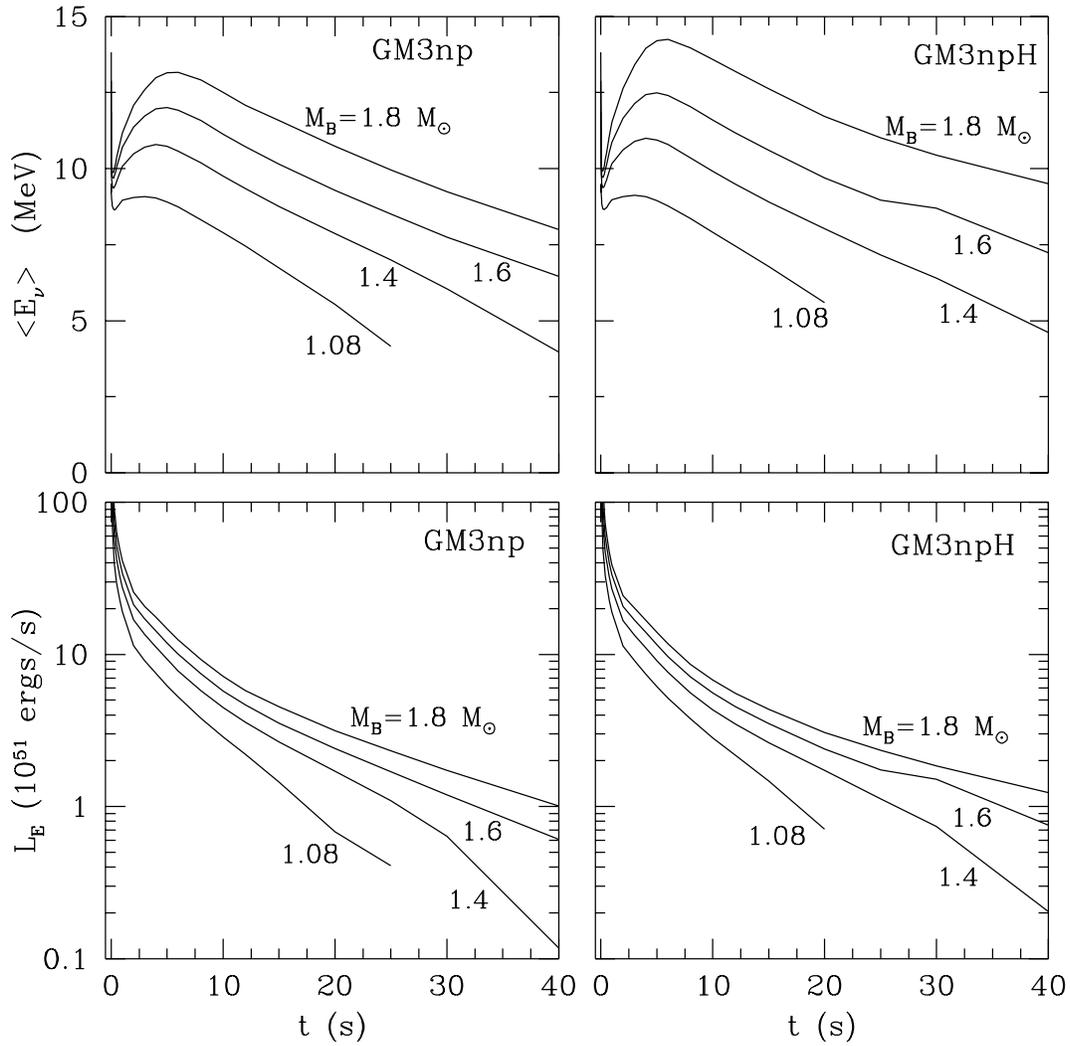}
\caption{The evolution of the average energy and total luminosity of
neutrinos in PNSs composed of baryons only (left panel)
and baryons and hyperons (right panel).  The figure is from
Ref. \cite{pons99}.}
\label{res_mass}
\end{figure}

\begin{figure}[hbt]
\begin{center}
\leavevmode
\includegraphics[scale=0.7]{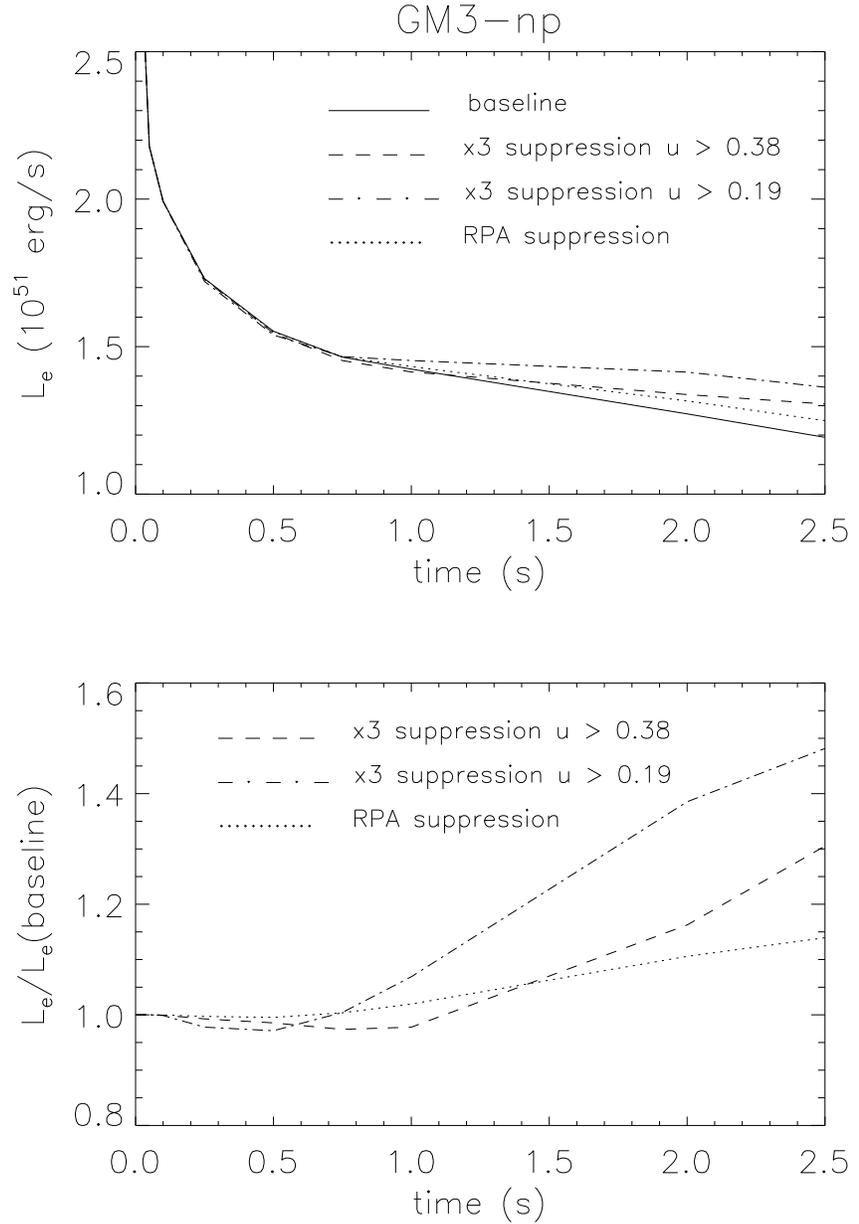}
\end{center}
\caption{ The upper panel shows the total emitted neutrino luminosity
for the evolution of a PNS using the opacities of
Ref. \cite{redd99a}.  Here $u=n_B/n_0$. The lower panel shows the ratio of the
luminosities obtained for models with correlation corrections to
the baseline (Hartree approximation) model.  This figure is from
Ref. \cite{pons99}.}
\label{lums1}
\end{figure}

\begin{figure}[hbt]
\begin{center}
\leavevmode
\includegraphics[scale=0.7]{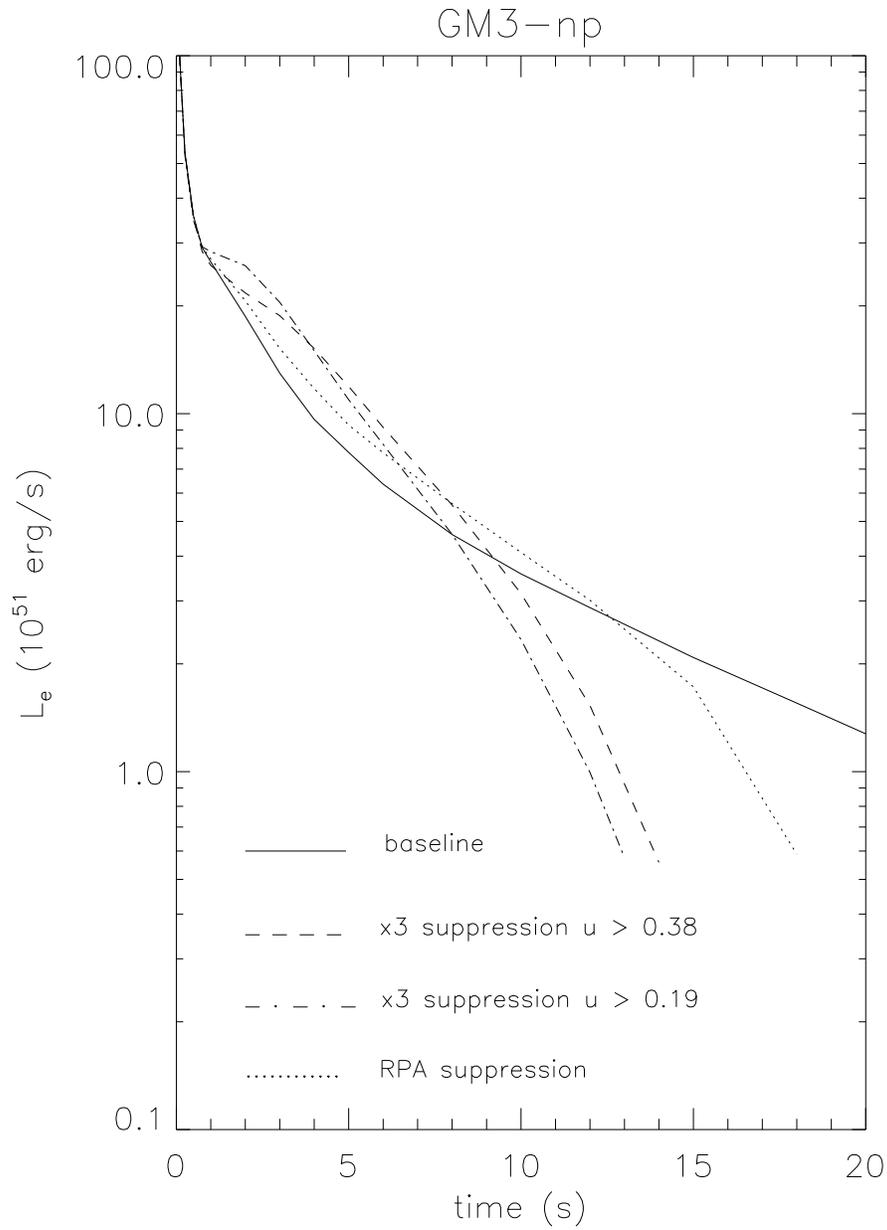}
\end{center}
\caption{Emitted neutrino luminosity in PNSs ($u=n_B/n_0$).  This
figure is from Ref. \cite{pons99}.}
\label{lums2}
\end{figure}

\begin{figure}[htb]
\begin{center}
\includegraphics[scale=0.7]{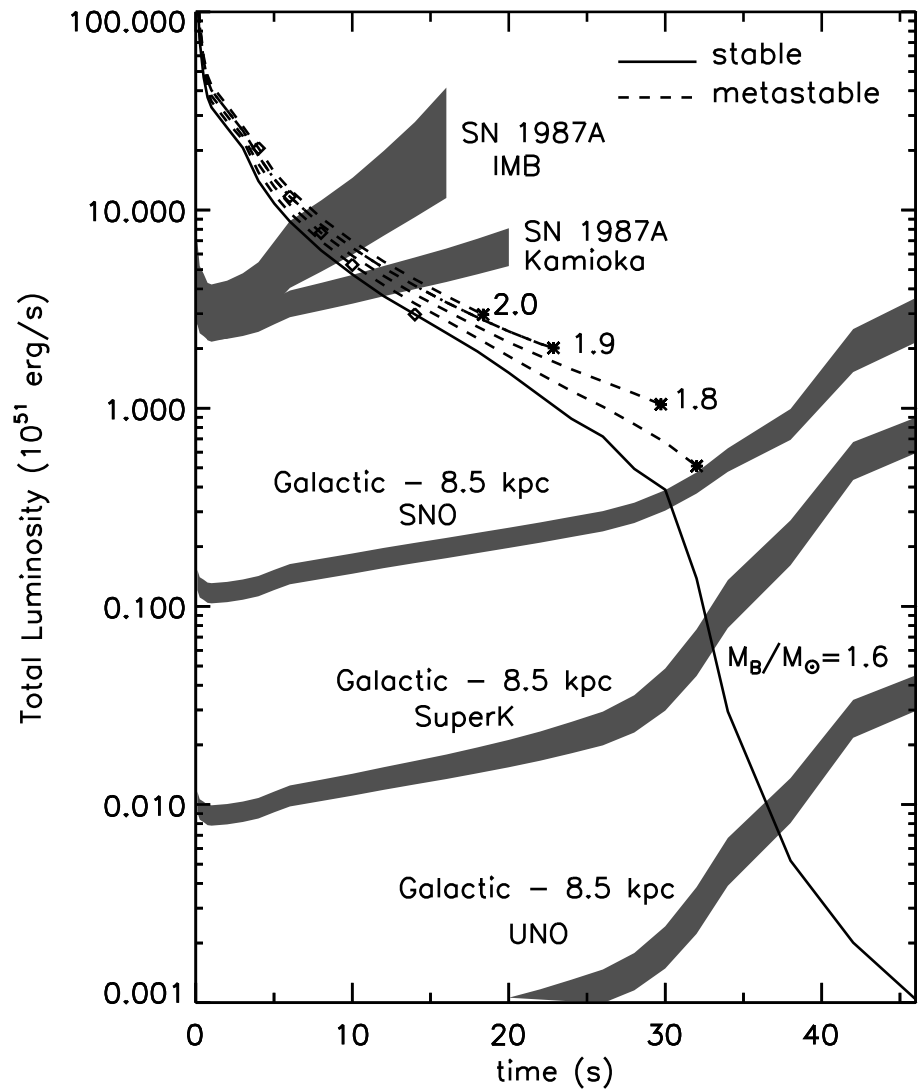}
\end{center}
\caption{The evolution of the total neutrino luminosity for $npQ$
PNSs.  Shaded bands illustrate the limiting
luminosities corresponding to a count rate of 0.2 Hz, assuming a
supernova distance of 50 kpc for IMB and Kamioka, and 8.5 kpc for SNO
and SuperK. The widths of the shaded regions represent uncertainties
in the average neutrino energy from the use of a diffusion scheme for
neutrino transport.  This figure is from Ref. \cite{Pon01b}.}
\label{fig:lum1}
\end{figure}

\begin{figure}[htb]
\begin{center}
\includegraphics[scale=0.7]{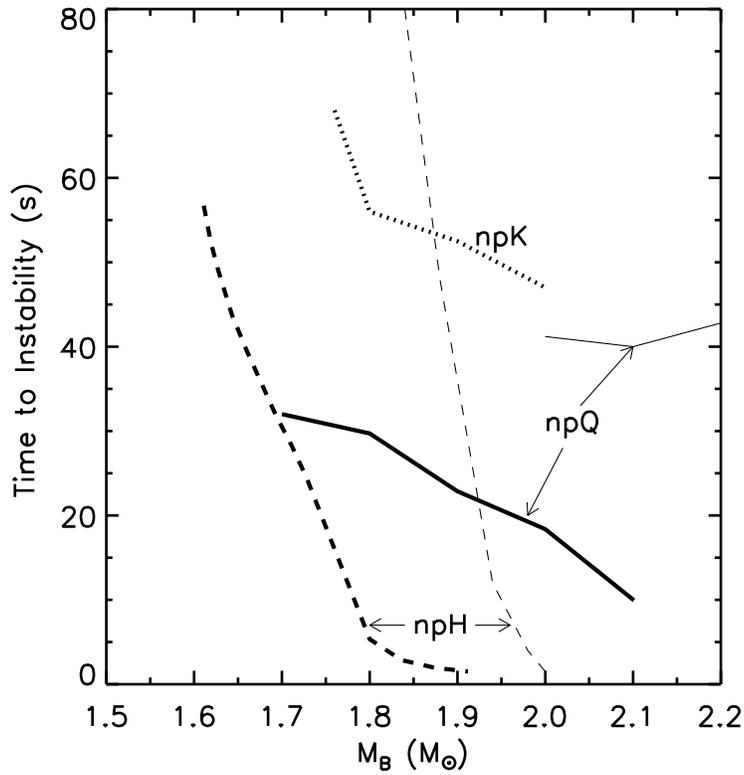}
\end{center}
\caption{Lifetimes of metastable stars versus the PNS
baryon mass $M_B$.  Thick lines denote cases in which the maximum
gravitational masses of cold, catalyzed stars are near 1.45 M$_\odot$,
which minimizes the metastability lifetimes.  The thin lines for the
$npQ$ and $npH$ cases are for EOSs with larger maximum gravitational
masses (1.85 and 1.55 M$_\odot$, respectively.)  This figure is from
Ref. \cite{Pon01b}.}
\label{fig:ttc}
\end{figure}

Neutrino signals from PNSs depends on many stellar properties,
including the mass; initial entropy, lepton fraction and density
profiles; and neutrino opacities.  In Figures ~\ref{res_mass} --
\ref{fig:ttc}, the dependence of neutrino emission on PNS
characteristics are shown from the detailed study of Pons et
al. \cite{pons99,Pon00a,Pon01b}.  The generic results (see
Figure~\ref{res_mass}) are that both $L_\nu$ and $<E_\nu>$ increase
with increasing mass~\cite{burr86,pons99}.  $<E_\nu>$ for all flavors
increases during the first 2-5 seconds of evolution, and then
decreases nearly linearly with time.  For times larger than about 10
seconds, and prior to the occurrence of neutrino transparency, the
$L_\nu$ decays exponentially with a time constant that is sensitive to
the high-density properties of matter. Significant variations in
neutrino emission occur beyond 10 seconds: $L_\nu$ is larger during
this time for stars with smaller radii and with the inclusion of
hyperons in the matter.  Finally, significant regions of the stars
appear to become convectively unstable during the evolution, as
several works have found \cite{convec}.

The main effect of the larger mean free paths produced by RPA
corrections \cite{redd99a, burr98,burr99a} is that the inner core
deleptonizes more quickly (see Figure~\ref{lums1}).  In turn, the
maxima in central temperature and entropy are reached on shorter
timescales. In addition, the faster increase in thermal pressure in
the core slows the compression associated with the deleptonization
stage, although after 10 s the net compressions of all models
converge.  The relatively large, early, changes in the central
thermodynamic variables do not, however, translate into similarly
large effects on observables such as $L_\nu$ and $<E_\nu>$, relative
to the baseline simulation.  It is especially important that at and
below nuclear density, the corrections due to correlations are
relatively small.  Since information from the inner core is
transmitted only by the neutrinos, the time scale to propagate any
high density effect to the neutrinosphere is the neutrino diffusion
time scale. Since the neutrinosphere is at a density approximately
$0.01n_0$, and large correlation corrections occur only above $n_0/3$
where nuclei disappear, correlation corrections have an effect at the
neutrinosphere only after 1.5 s.  However, the corrections are still
very important during the longer-term cooling stage (see
Figure~\ref{lums2}), and result in a more rapid onset of neutrino
transparency compared to the Hartree results.

\subsubsection{Neutrino Signals in Terrestrial Detectors}

A comparison of the signals observable with different detectors is
shown in Figure \ref{fig:lum1}, which displays $L_\nu$ as a function
of baryon mass $M_B$ for stars containing quarks in their cores.  In
the absence of accretion, $M_B$ remains constant during the evolution,
while the gravitational mass $M_G$ decreases.  The two upper shaded
bands correspond to estimated SN 1987A (50 kpc distance) detection
limits with KII and IMB, and the lower bands correspond to estimated
detection limits in SNO, SuperK, and UNO, for a Galactic supernova
(8.5 kpc distance).  The detection limits have been set to a count
rate $dN/dt=0.2$ Hz \cite{Pon00a}.  It is possible that this limit is
too conservative and could be lowered with identifiable backgrounds
and knowledge of the direction of the signal.  The width of the bands
represents the uncertainty in $<E_{\bar\nu_e}>$ due to the diffusion
approximation \cite{pons99,Pon00a,Pon01b}.  It appears possible to
distinguish between stable and metastable stars, since the
luminosities when metastability is reached are always above
conservative detection limits.

\subsubsection{Metastable Proto-Neutron Stars}

Proto-neutron stars in which strangeness appears following
deleptonization can be metastable if their masses are large enough.
One interesting diagnostic that could shed light on the internal
composition of neutron stars would be the abrupt cessation of the
neutrino signal.  This would be in contrast to a normal star of
similar mass for which the signal continues to fall until it is
obscured by the background.  In Figure~\ref{fig:ttc} the lifetimes for
stars containing hyperons ($npH$), kaons ($npK$) and quarks ($npQ$)
are compared \cite{Pon00a}.  In all cases, the larger the mass, the
shorter the lifetime.  For the kaon and quark PNSs, however, the
collapse is delayed until the final stage of the Kelvin-Helmholtz
epoch, while this is not necessarily the case for hyperon-rich stars.
In addition, there is a much stronger mass dependence of the lifetimes
for the hyperon case.

Clearly, the observation of a single case of metastability, and the
determination of the metastability time alone, will not necessarily
permit one to distinguish among the various possibilities. Only if the
metastability time is less than 10--15 s, could one decide on this
basis that the star's composition was that of $npH$ matter.  However,
as in the case of SN 1987A, independent estimates of $M_B$ might be
available \cite{THM90BB95}.  In addition, the observation of two or
more metastable neutron stars might permit one to differentiate among
these models.

\section{OUTLOOK}

On the early universe front, an important issue is simply whether or
not light sterile neutrinos exist. If they exist, then the details of
light element synthesis could be interestingly different from that of
standard Big Bang Nucleosynthesis, because of the neutrino asymmetry
amplification phenomenon. One must look forward to crucial
experimental results from SNO, MiniBOONe and the long baseline
facilities. Naturally, further observational study of the actual light
element abundances is a central concern. Precision cosmic microwave
background anisotropy measurements will also be important as an
independent probe of the baryon to photon ratio, and of the expansion
rate of the early universe. An interesting theoretical issue currently
under examination is that of inhomogenous neutrino asymmetry creation
\cite{inhom1,inhom2}.

The outlook from the supernova perspective hinges on technical
advances in handling multigroup, general relativistic, Boltzmann
neutrino transport.  The advent of next-generation neutrino detectors
such as Super-Kamiokande and the Sudbury Neutrino Observatory promises
thousands of neutrino events in the next Galactic supernova.  These
will provide crucial diagnostics for the supernova mechanism,
important limits on the released binding energy and the remnant mass,
and critical clues concerning the composition of high density matter.
Research in this area will ascertain the extent to which neutrino
transport is instrumental in making a supernova explode.  Other
bonuses include the elucidation of the possible role of supernovae and
neutrinos in $r-$process nucleosynthesis.

The main issues that emerge from PNS studies concern the metastability
and subsequent collapse to a black hole of a PNS containing quark
matter, or other types of matter including hyperons or a Bose
condensate, which could be observable in the $\nu$ signal.  However,
discriminating among various compositions may require more than one
such observation.  This highlights the need for breakthroughs in
lattice simulations of QCD at finite baryon density in order to
unambiguously determine the EOS of high density matter.  In the
meantime, intriguing possible extensions of supernova and PNS
simulations with $npQ$ and $npK$ matter include the consideration of
heterogenoeus structures and quark matter superfluidity \cite{CR00}.

\paragraph{Acknowledgements}  Research support from DOE grants 
FG02-88ER-40388 (for MP) and FG02-87ER-40317 and the J.S. Guggenheim
Foundation (for JML) is gratefully acknowledged.  MP and JML thank
Sanjay Reddy, Jose Pons, and Andrew Steiner for benificial
collaborations.  RFS thanks Adam Burrows for valuable collaborations.
His work is supported in part by NSF grant PHY-9900544.  RRV would
like to thank Nicole Bell, Pasquale Di Bari, Robert Foot, Keith Lee,
Mark Thomson and Yvonne Wong for exciting and fruitful collaborations
on neutrino cosmology.  He is supported by the Australian Research
Council and the University of Melbourne.

\newpage

\newpage

\newpage

\newpage
\newpage

\newpage

\newpage

\newpage

\newpage

\newpage

\newpage

\newpage

\newpage

\newpage

\end{document}